\documentclass[aps,pra,twocolumn,superscriptaddress,groupedaddress]{revtex4}  
\usepackage{graphicx}  
\usepackage{dcolumn}   
\usepackage{bm}        
\usepackage{amssymb}
\usepackage{amsmath} 
\usepackage[caption=false]{subfig}
\usepackage{float} 
\usepackage{lipsum} 
\usepackage{color}
\usepackage{subfig}
\usepackage{qcircuit}
\usepackage{setspace}

\usepackage{array}

\newcolumntype{L}[1]{>{\raggedright\arraybackslash}m{#1}}
\newcolumntype{C}[1]{>{\centering\arraybackslash}m{#1}}
\newcolumntype{R}[1]{>{\raggedleft\arraybackslash}m{#1}}

\newcommand{\be}{\begin{equation}}
\newcommand{\ee}{\end{equation}}
\newcommand{\ba}{\begin{eqnarray}}
\newcommand{\ea}{\end{eqnarray}}

\def\ben{\begin{equation}}
\def\een{\end{equation}}

\def\bea{\begin{eqnarray}}
\def\eea{\end{eqnarray}}

\begin{document}
\newcount\hour \newcount\minute
\hour=\time  \divide \hour by 60
\minute=\time
\loop \ifnum \minute > 59 \advance \minute by -60 \repeat
\def\nowtwelve{\ifnum \hour<13 \number\hour:
                      \ifnum \minute<10 0\fi
                      \number\minute
                      \ifnum \hour<12 \ A.M.\else \ P.M.\fi
	 \else \advance \hour by -12 \number\hour:
                      \ifnum \minute<10 0\fi
                      \number\minute \ P.M.\fi}
\def\nowtwentyfour{\ifnum \hour<10 0\fi
		\number\hour:
         	\ifnum \minute<10 0\fi
         	\number\minute}
\def \now {\nowtwelve}

\title{An Experimental Study of Shor's Factoring Algorithm on IBM Q}

\author{Mirko Amico}
 \affiliation{The Graduate School and University Center, The City University of New York, New York, NY 10016, USA}
\author{Zain H. Saleem}
\affiliation{Theoretical Research Institute of Pakistan Academy of Sciences, Islamabad 44000, Pakistan}
\author{Muir Kumph}
\affiliation{IBM T.J. Watson Research Center, Yorktown Heights, NY 10598, USA}

\begin{abstract}

We study the results of a compiled version of Shor's factoring algorithm on the \textit{ibmqx5} superconducting chip, for the particular case of $N=15$, $21$ and $35$. The semi-classical quantum Fourier transform is used to implement the algorithm with only a small number of physical qubits and the circuits are designed to reduce the number of gates to the minimum. We use the square of the statistical overlap to give a quantitative measure of the similarity between the experimentally obtained distribution of phases and the predicted theoretical distribution one for different values of the period. This allows us to assign a period to the experimental data without the use of the continued fraction algorithm. A quantitative estimate of the error in our assignment of the period is then given by the overlap coefficient.     
\end{abstract}

\preprint{}

\maketitle

\section{Introduction}

Shor's factoring algorithm \cite{shor} is a well known example of a quantum algorithm outperforming the best known classical algorithm. Experimental implementation of the algorithm with physical qubits however remains a challenge because of the errors introduced by the large number of qubits and gates required to execute the algorithm. In this article we provide a proof-of-principle demonstration of a compiled version of Shor's factoring algorithm to factor the numbers $N=15$, $21$ and $35$ using five, six and seven superconducting qubits, respectively. Similar experiments have been done on setups like NMR \cite{vandersypen}, trapped ions \cite{monz}, photons \cite{lu, lanyon, lopez}, photonic chips \cite{politi} and superconducting qubits \cite{lucero, coles}. However, with the exception of Refs. \cite{monz, lopez}, all these realizations involve an oversimplified version of the algorithm which is equivalent to coin flipping \cite{smolin} and no quantum hardware is needed to obtain the same results. 

In our implementation, classical processing is used alongside quantum computation to overcome the lack of key-functions of the device used. Furthermore, the number of physical qubits and the circuit depth are reduced to the minimum in order to minimize the effects of noise. The data is presented as estimates of the probability distribution of the values returned by the period register.  While obtaining the probability distributions of the period register requires running the algorithm many times, as opposed to just once with the original continued fraction expansion, this allows the performance of quantum computers running this algorithm to be more directly evaluated. To measure the success of the experiments in different ways, the results are analyzed in both a qualitative way, with probability plots \cite{kleiner}, and a quantitative way, with the square of statistical overlap (SSO) \cite{chiaverini}. Probability plots are a useful tool to visualize differences between probability distributions, while the SSO provides a quantitative measure of their similarity. Using the the overlap coefficient (see below), we can also use the SSO to assign a period to the experimental data, avoiding the continued fraction algorithm which does not work for such low number of qubits. Also, the overlap coefficient (OVL) gives an estimate of the probability that the experiment succeeded.  The results of the experiments are in good agreement with the theory for $N=15$ and $21$. However, the experiment succeeded for $N=35$ only about 14\% of the time, where the cumulative errors coming from the high number of two-qubit gates became too large.

The article is organized in the following way. A brief overview of Shor's factoring algorithm is given in Sec. \ref{shor}. Sec. \ref{hardware} describes the hardware used for the experiment. In Sec. \ref{exp} the implementation of the factoring experiment for $N=15$, $21$ and $35$, respectively, is described. The results obtained from running the algorithm on the \textit{ibmqx5} quantum processor are analyzed and discussed in Sec. \ref{data}. Conclusions follow in Sec. \ref{conclusions}.

\section{Overview of Shor's factoring algorithm}
\label{shor}
The factoring algorithm invented by P. Shor \cite{shor} relies on the relation between the problem of factoring and the problem of order finding, for which a quantum speed-up exists. In fact, finding the prime factors of a number $N$ is equivalent to finding the exponent $x$ for which the function $a^x \text{mod} N =1$, where $a$ is an integer smaller than $N$ picked at random. Such exponent is called the order, or period, of $a$. Let us briefly review the quantum part of the algorithm before diving into the details of the experiment.
Two quantum registers are needed for the computation. One register is used to store the value of the period, called period register, and the other to store the results of the computation, called computational register. The size of both registers depends on the number $N$ to be factored. In particular, the period register should have a number of qubits $n_p$ in the interval $\text{log}_2(N^2) \lesssim n_p \lesssim \text{log}_2(2N^2)$ and the computational register should be large enough to be able to represent the number $N-1$, resulting from the modular exponentiation function (MEF) $a^x  \text{mod} N$, thus requiring $n_q = \text{log}_2 N$ qubits.

At the beginning of the quantum algorithm, the two registers are initialized to the state $\lvert 00...0 \rangle_p \lvert 00...1 \rangle_q $, where the subscripts $p$ and $q$ denote the period register and the computational register, respectively. The period register stores all the possible values of the exponent $x$, which will give an estimate of the period, by creating a uniform superposition of all possible bit-strings through Hadamard gates on all qubits $\frac{1}{\sqrt{Q}} \sum_{x=0}^{Q-1} \lvert x \rangle_p$, where $Q = 2^{n_p}$. While the computational register stores the results of the MEF, $a^x  \text{mod} N$. After the first step, the two registers are in the state $\frac{1}{\sqrt{Q-1}} \sum_{x=0}^{Q-1} \lvert x \rangle_p \lvert a^x \text{mod} N \rangle_q$. Then, the quantum Fourier transform (QFT) is applied to the period register so that $\lvert x \rangle_p \rightarrow \frac{1}{\sqrt{Q}} \sum_{s=0}^{Q-1} e^{\frac{2\pi i s x}{Q}}\lvert s \rangle_p$. As a result of the QFT, interference between all the possible states occurs. If the period register is then measured, a value of the phase $s$ is measured with probability $\text{P}(s) = \frac{1}{Q} \sum_{s=0}^{Q-1} \lvert e^{\frac{2\pi i s x}{Q}} \rvert^2$. Rewriting $x$ in terms of the period $r$ as $x = x_0 + d r $, where $x_0$ and $d$ are integers, the probability of an outcome $s$ can be written as $\text{P}(s) = \frac{1}{Q} \lvert e^{\frac{2\pi i x_0}{Q}} \rvert^2 \sum_{s=0}^{Q-1}  \lvert e^{\frac{2\pi i s d r}{Q}} \rvert^2$. Clearly, a value of $s$ such that $\frac{s}{Q}=\frac{c}{r}$, where $c$ is an integer, will be observed with high probability.

The final part of the algorithm involves classical processing of the measurement obtained in the quantum part. The value of the period $r$ can be found from the fraction $\frac{s}{Q}$ by using the continued fraction algorithm. Or as done in this paper, by running the algorithm many times to get a direct estimate of the probability distribution of the values for the period register.  A comparison between the measured probability distribution and the theoretically predicted distribution for the period $r$ can be made using the SSO and the best fit gives the most likely period.  If the period $r$ calculated in this way is odd or $r=0$, the algorithm fails and one restarts by picking a different base $a$. If $r$ is even, $(a^r - 1) \text{mod} N$ can be factored into $\left( a^{\frac{r}{2}} - 1\right)\left( a^{\frac{r}{2}} + 1\right) \, \text{mod} N$. The final step is to check if $(a^{\frac{r}{2}} + 1)  \text{mod} N$ has a common divisor with $N$ by checking that gcd$\left(a^{\frac{r}{2}} + 1,N \right) \neq 1$. If that's true, then the two factors of $N$ are gcd$\left(a^{\frac{r}{2}} + 1,N \right)$ and gcd$\left(a^{\frac{r}{2}} - 1,N \right)$.

As mentioned earlier, the execution of this version of the algorithm requires $n_q=\text{log}_2 \left( N \right)$ qubits in the computational register to perform the modular exponentiation and at least another $n_p = 2\text{log}_2 \left( N \right)$ qubits in the period register to perform the QFT. Thus the complete algorithm requires a total number of $3\text{log}_2 \left( N \right)$ qubits. Even the factoring of a number as small as $N=15$ needs $12$ qubits in the input register to execute this algorithm, which is still a challenge for today's physical realizations of quantum computers. However, Kitaev \cite{kitaev} observed that for the purpose of algorithms like Shor's, where one doesn't need the information on the relative phase of the output states but only their measured probability amplitudes, one can replace the fully coherent quantum Fourier transform with the semi-classical quantum Fourier transform (sc-QFT). In the sc-QFT, one of the qubits of the period register is measured each time. The result of the measurement of the qubit is then used to determine the type of measurement on the next one. This enables the replacement of the $2\text{log}_2 \left( N \right)$ qubits of the period register with a single qubit measured multiple times. For the case of factoring $N=15$, Kitaev's approach reduces the total number of qubits required to $n=5$ and for the case of $N=21$ and $N=35$ to $n=6$ and $n=7$, respectively, which are small enough numbers for the presently available hardware to handle. This decrease in the system size, however, comes with the drawback of requiring in-sequence single-qubit readout and state re-initialization together with feed-forward of gate settings based on previous measurement results. The implementation of the sc-QFT has been described in \cite{griffiths, chiaverini} and realized in \cite{monz}. At present the IBM quantum computer doesn't perform in-sequence single qubit read out and qubit re-initialization. Below, we provide a procedure for going around this hurdle to implement the sc-QFT on the IBM Q device.

\section{Hardware}
\label{hardware}
We use the IBM {\it ibmqx5} chip with sixteen superconducting qubits to implement our experiments for factoring the numbers $N=15$, $21$ and $35$. 
The qubits are distributed on the plane, as two adjacent arrays of eight qubits each with couplings shown in Fig. \ref{fig:ibmqx5_coupling}.

\begin{figure}[]
	\includegraphics[width = 3.5in, height = 0.8 in]{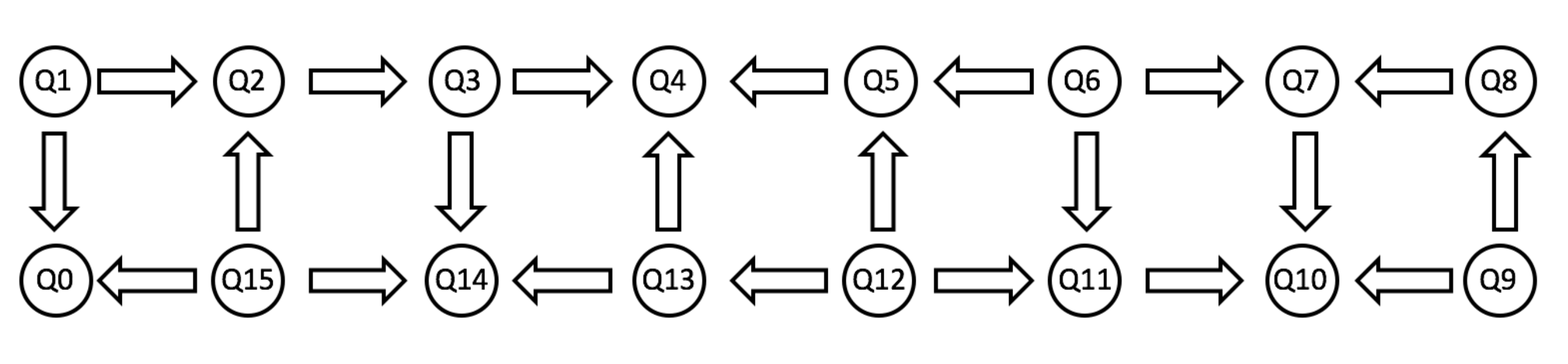} \\
	\caption{Coupling map between the 16 qubits of the \textit{ibmqx5} device. The arrows indicate the possible CNOT gates between pairs of qubits. In particular, the arrow starts from the control qubit and points to the target qubit.}
\label{fig:ibmqx5_coupling}
\end{figure}

The qubits' relaxation time $T_1$ ranges from $25 \sim 60$ $\mu$s and their dephasing time $T_2$ from $20 \sim 100$ $\mu$s. The single-qubit gates have a high fidelity, measured at $\sim 99.8 \% $ at the time of the experiment. The multi-qubit gate fidelity was measured around $95\% - 98 \%$ depending on the pairs of qubits considered. All gate errors are measured using simultaneous randomized benchmarking. Another source of error comes from the read-out of the states of the qubits, which amounts to roughly an error of $5 \%$. Using these parameters, the effects of noise can be incorporated in the simulation, obtaining a more accurate prediction for the output of the device.

\section{Experiment}
\label{exp}

\begin{figure}[h]
\includegraphics[width = 3.4in, height = 0.7 in]{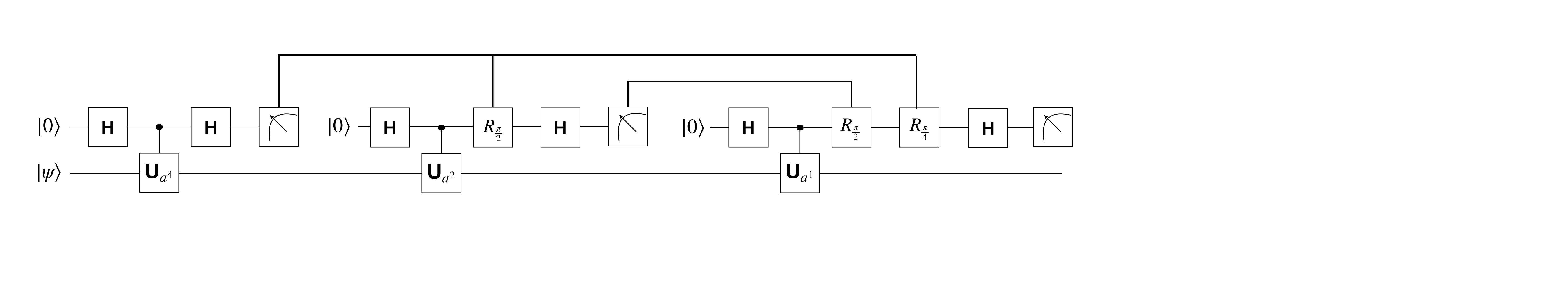}
	\caption{Circuit for factoring $N=15$, $21$ and $35$ implemented using the scheme shown in Ref. \cite{monz}. The first register (top), the period register, stores the estimation of the period, this can be done by using only one qubit through the sc-QFT. The second register (bottom), the computational register, stores the outcomes of the modular exponentiation function $a^x \text{mod} N$ computed through the controlled-$U_{a^x}$ gate. The specific circuits for $U_{a^x}$ used in factoring $N=15,21$ and $35$ for a specific base $a$ are shown in Appendix \ref{app_A}. }
\label{fig:full_circuit}
\end{figure}

Following the example given in \cite{monz}, we implement the quantum part of Shor's factoring algorithm using the circuit depicted in Fig. \ref{fig:full_circuit}. As can be seen in the circuit diagram in Fig. \ref{fig:full_circuit}, rotations of the control qubit depend on the outcome of each of its measurement in the previous steps. Since the \textit{ibmqx5} chip does not allow for qubit reset and conditional operation based on measurements, which are required to implement the sc-QFT suggested by Kitaev, we implement the algorithm as three separate quantum circuits as shown in Fig. \ref{fig:full_circuit_broken}. 

In the first circuit, the system is initialized in the state $\lvert 0 \rangle_p \lvert 0...01 \rangle_q$ and the first bit, $b_0$, encoding the value of the period is measured at the end. In the second circuit, the initial state $\lvert 0 \rangle_p \lvert \psi_{b_0} \rangle_q$ is prepared. Different states $\lvert \psi_{b_0}$ are prepared depending on the value of $b_0$ measured in the previous circuit. Rotation gates on the period register are also inserted conditional on the value of $b_0$ before measuring the second bit encoding the value of the period, $b_1$. In the third circuit, depending on the values of $b_0$ and $b_1$, the qubit registers are initialized to $\lvert 0 \rangle_p \lvert \psi_{b_0 b_1} \rangle_q$ and rotation gates are inserted before the measurement of $b_2$. The possible quantum states of the computational register can be computed classically for the full algorithm, conditional on the measurement results of the period register. This is just the result of successive modular exponentiation. At the beginning of each circuit, except the first one, there are two possible states of the computational register that have to be prepared depending on the value of the period register measured in the previous stage. If the measurement of the period register gives $0$, then the computational register is prepared to the state $\lvert \psi_{0} \rangle_q = \lvert \psi \rangle_q + Z U_{a^x}  \lvert \psi \rangle_q $. If the period register gives $1$, the computational register is initialized to $\lvert \psi_{1} \rangle_q = \lvert \psi \rangle_q - Z U_{a^x}  \lvert \psi \rangle_q $. This means that for an implementation with $m$ stages, a superposition of $2^m$ product states have to be prepared. However, the state at the $m$-th stage is, at worst, the result of $m-1$ modular exponentiations. Thus, breaking the circuit in this way, only adds an extra number of gates which is polynomial in the number of stages $m$: $1+2+3+...+ m = \frac{m(m+1)}{2}$, due to the gates needed for the state initialization at each stage. This retains the scalability of the implementation given in Ref. \cite{monz}.

In the following experiments, we limit ourselves to the choice of one, or two, bases $a$ to avoid redundancy. We specifically choose a non-trivial base (in the sense of \cite{smolin}) for which a working quantum processor is needed to find the results. One could adopt the same approach to treat any such non-trivial bases. To understand what happens in the case of a trivial base, consider factoring $N=15$. The possible periods $r$ for any of the bases $a$ are all powers of two. This means that any even value of the phase $s$ measured from the period register will give a fraction $\frac{s}{Q}$ proportional to $\frac{1}{r}$ which always allows one to find the period. In facts, by analyzing the state of the quantum registers along the circuit, it is possible to see that no quantum interference happens between the states in the computational register. Therefore, in this case the correct results can be obtained regardless of the quality of the entangling gates of the device, as long as one can entangle the period register with the computational register. To show that the quantum processor \textit{ibmqx5} is giving us the correct answer by exploiting quantum interference it is sufficient to run the experiment for one of the possible bases. This is in turn is related to the quality of the entangling gates and the noise of the device. Thus, the ability to factor higher and higher $N$ using a non-trivial base (one which has a period that is not a power of 2) gives a benchmark of the performance of the device.

In the experiment for the $N=15$ case, five input qubits are required. One qubit initialized to $\lvert 0 \rangle_p$ for the period register, acting as a control qubit, and all other qubits initialized to the state $\lvert \psi \rangle =  \lvert 0 0 0 1 \rangle_q$ belonging to the computational register. Alongside the quantum registers, we also need a three-bit classical register to store the results of the measurement of the control qubit, which encodes the value of the period.

\begin{figure}[]
\includegraphics[width = 3.06in, height = 2.7 in]{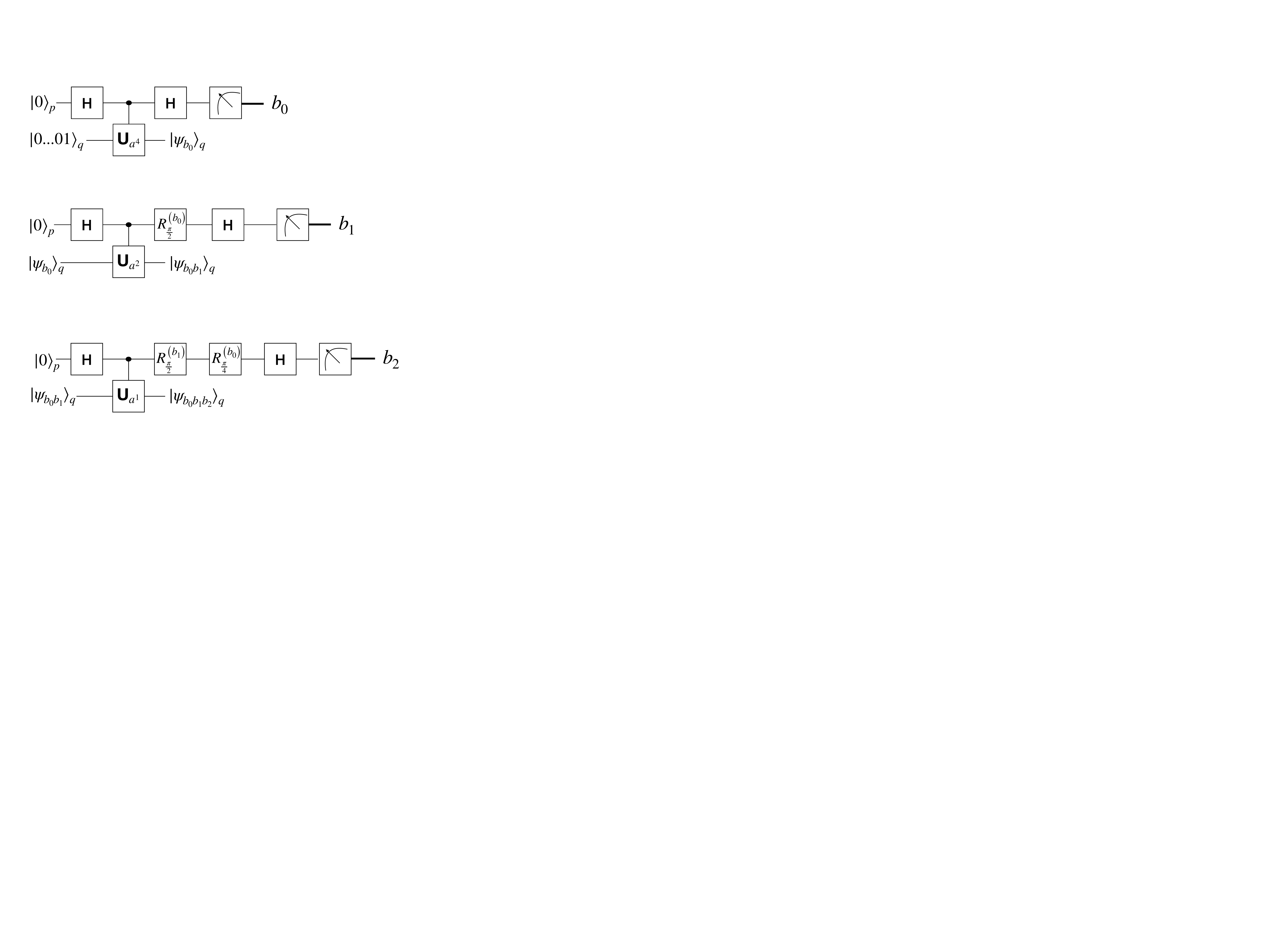}
	\caption{Circuits used in the experimental implementation of Shor's algorithm on \textit{ibmqx5}. The circuit of Fig. \ref{fig:full_circuit} is divided in three separate parts. Each circuit containing a stage of modular exponentiation and a measurement of the period register. The different circuits are joined using a classical algorithm which computes the quantum state of the computational register at the end of the previous circuit and feeds it as input to the next circuit. The classical algorithm also adds the right rotation gates on the period qubit in each successive circuit, based on the results of previous measurements. }
\label{fig:full_circuit_broken}
\end{figure}

The case $N=15$ is the simplest possible case and it does not provide an example where quantum interference between the states of the computational register brings an advantage to the computation. For this reason, we attempt to factor the second smallest number which is a product of two primes, $N=21$. In this case, there are bases $a$ for which the period is not a power of two, thus constructive quantum interference between states in the computational register is needed to increase the likelihood of finding the correct result. An example of such case was first demonstrated in Ref. \cite{lopez}.

We implement an algorithm for factoring $N=21$ with base $a=2$ using three bits of precision for the estimation of the phase which encodes the period. In this case the quantum register is composed of five qubits in the computational register and one qubit in the period register. We adopt the same methodology used previously, breaking each stage of the modular exponentiation and manually feeding the output of each section as input to the next. This means that the circuit will have three stages of modular exponentiation, where a single bit of the phase which encodes the period is estimated at each stage (details in Appendix \ref{app_A}). Therefore, the circuit looks like the one in Fig. \ref{fig:full_circuit_broken}. The modular exponentiation circuit are specifically designed to calculate $a^{x} \, \text{mod} \, 21$, where we choose the base $a=2$. This base has periods $r=6$, thus $1/r$ cannot easily be represented in binary. Therefore, the accuracy of the estimation of the period depends on the number of bits used for the phase estimation.

The same method is applied to factor $N=35$ with base $a=4$. In this case we need six qubit in the computational register and one qubit in the period register. As in the case of $N=21$, the period of $4^{x} \, \text{mod} \, 35$ is $r=6$, therefore $1/r$ cannot be easily represented in binary. As a result of running the quantum algorithm we obtain a probability distribution for the estimated phase $s$ which is peaked around the multiples of $1/r$. We use a three-bit register for the estimation of the phase which encodes the period. Again, the circuit for running the algorithm is realized as shown in Fig. \ref{fig:full_circuit}, each stage, estimating one bit of the phase, is implemented separately and then joined through a classical algorithm. The individual circuits which compute the MEF at the different stages can be found in Appendix \ref{app_A}.

\section{Results and Data Analysis}
\label{data}

Figs. \ref{fig:hist_2mod15}, \ref{fig:hist_11mod15}, \ref{fig:hist_2mod21_2} and \ref{hist_4mod35} show the results obtained running the quantum part of the factoring algorithm on the \textit{ibmqx5} superconducting device. Depicted are the experimental relative probabilities found (in blue) side by side with the expectation values which can be computed theoretically (in green) for each value of the estimated phase $s$ for the bases $a$ used. The algorithm was run 1000 times for each base.

\begin{figure}[]
	\subfloat[]{\label{fig:hist_2mod15}\includegraphics[width = 3.2in, height = 2.0 in]{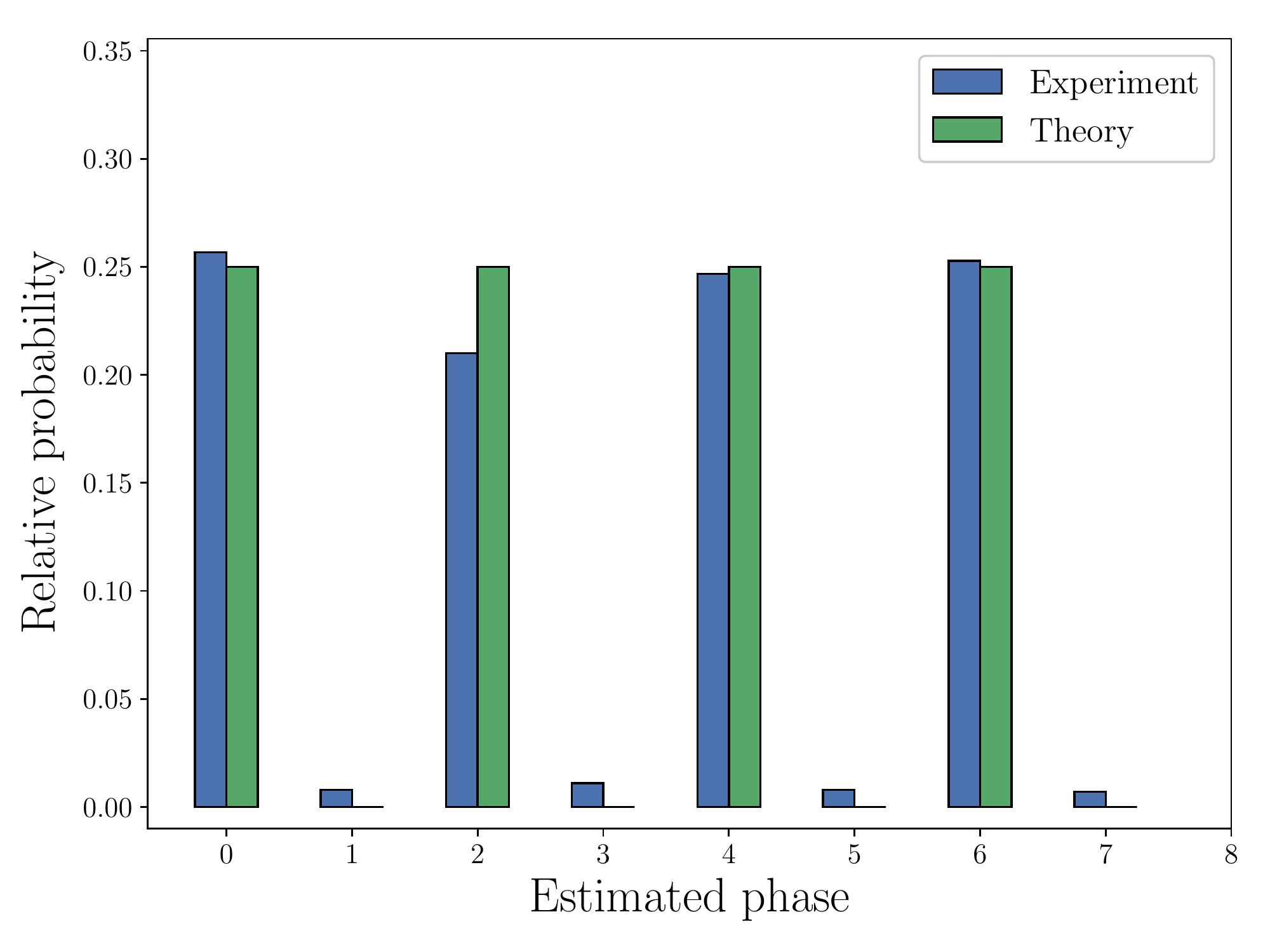}} \\
	\subfloat[]{\label{fig:pp_plot_2_15}\includegraphics[width = 3.6in, height = 2.0 in]{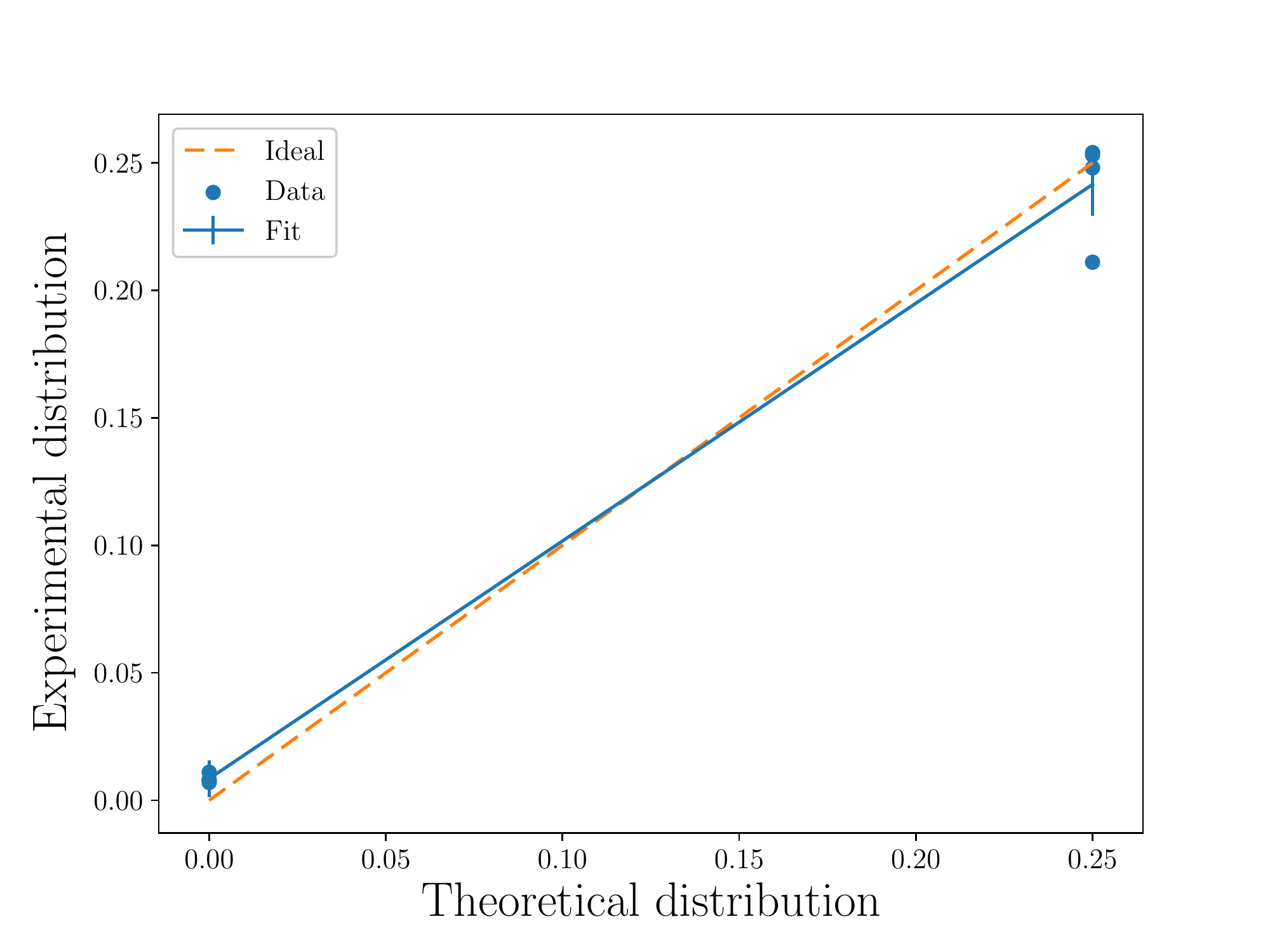}}\\
	\subfloat[]{\label{fig:sso_2_15}\includegraphics[width = 3.2in, height = 2.0 in]{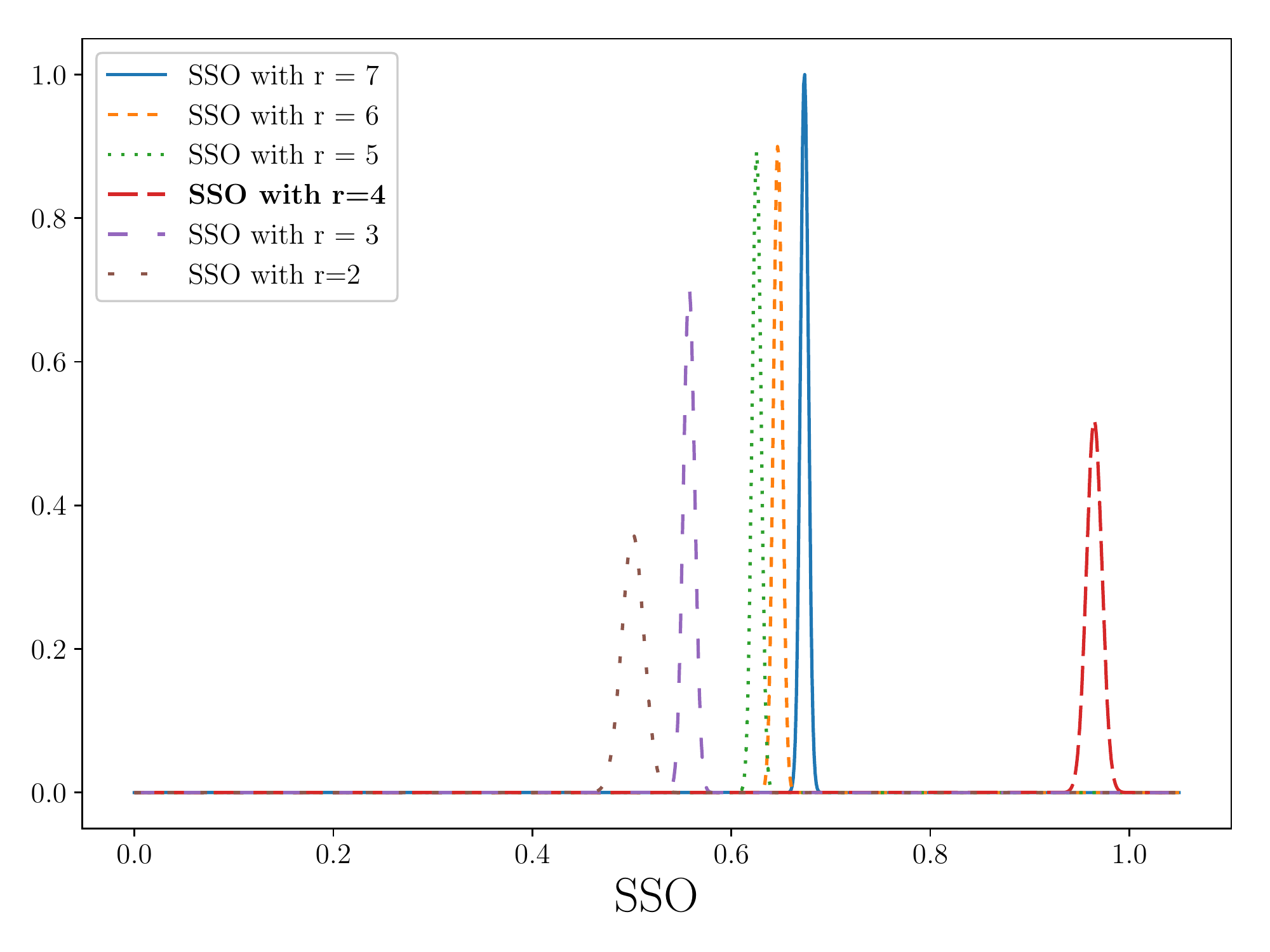}}  
	\caption{(a) Probability of finding a given phase for $N=15$ with base $a=2$, and (b) probability plot of the theoretical distribution and the experimental distribution for $r=4$. The experimental distribution is depicted through the collection of data and a fit of the data. (c) SSO of the experimental data with the theoretical probability distribution corresponding to all possible values of the period $r$. }
\label{histo_2_15}
\end{figure}

\begin{figure}[]
	\subfloat[]{\label{fig:hist_11mod15}\includegraphics[width = 3.2in, height = 2.0 in]{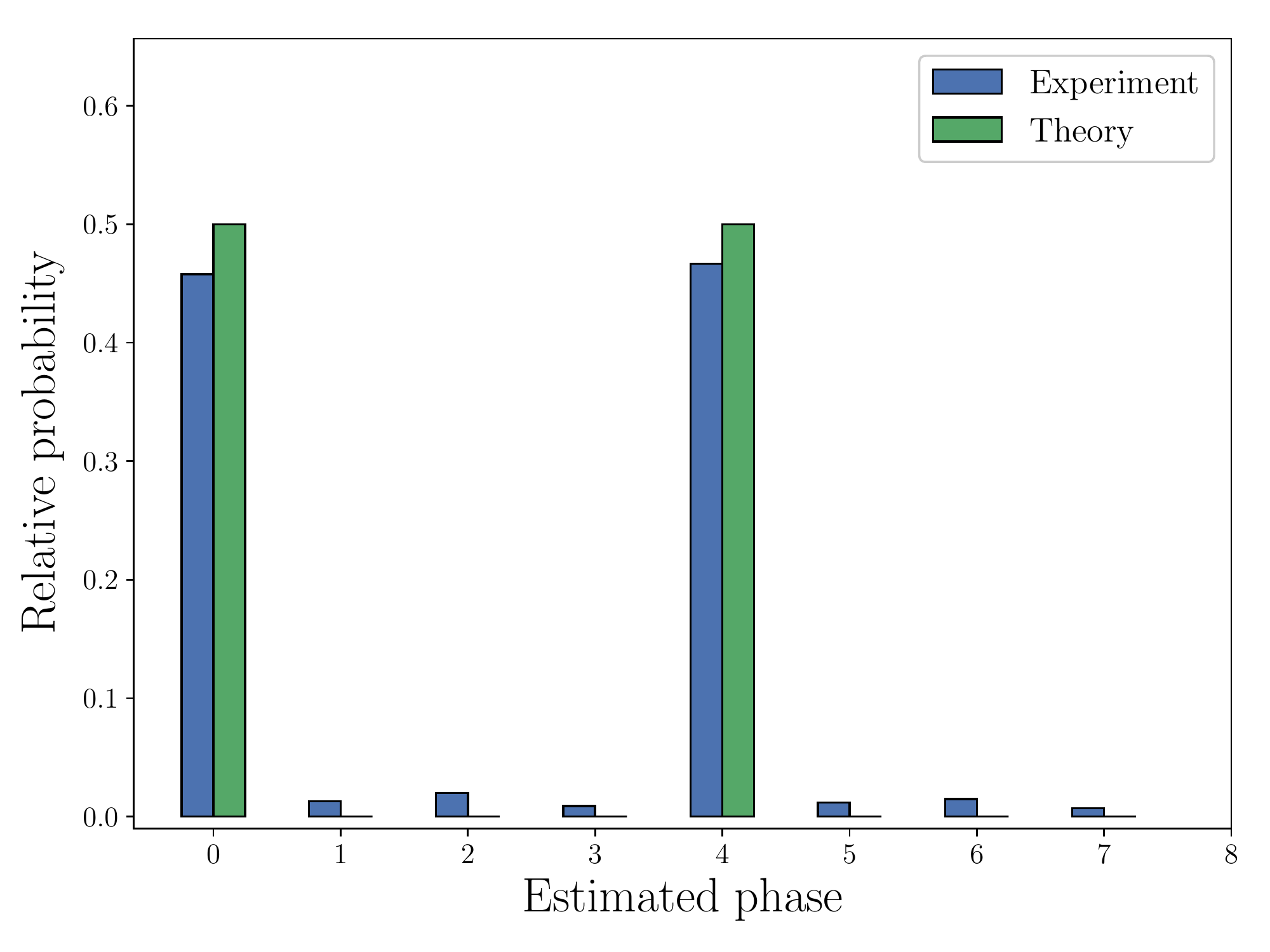}} \\
	\subfloat[]{\label{fig:pp_plot_11_15}\includegraphics[width = 3.6in, height = 2.0 in]{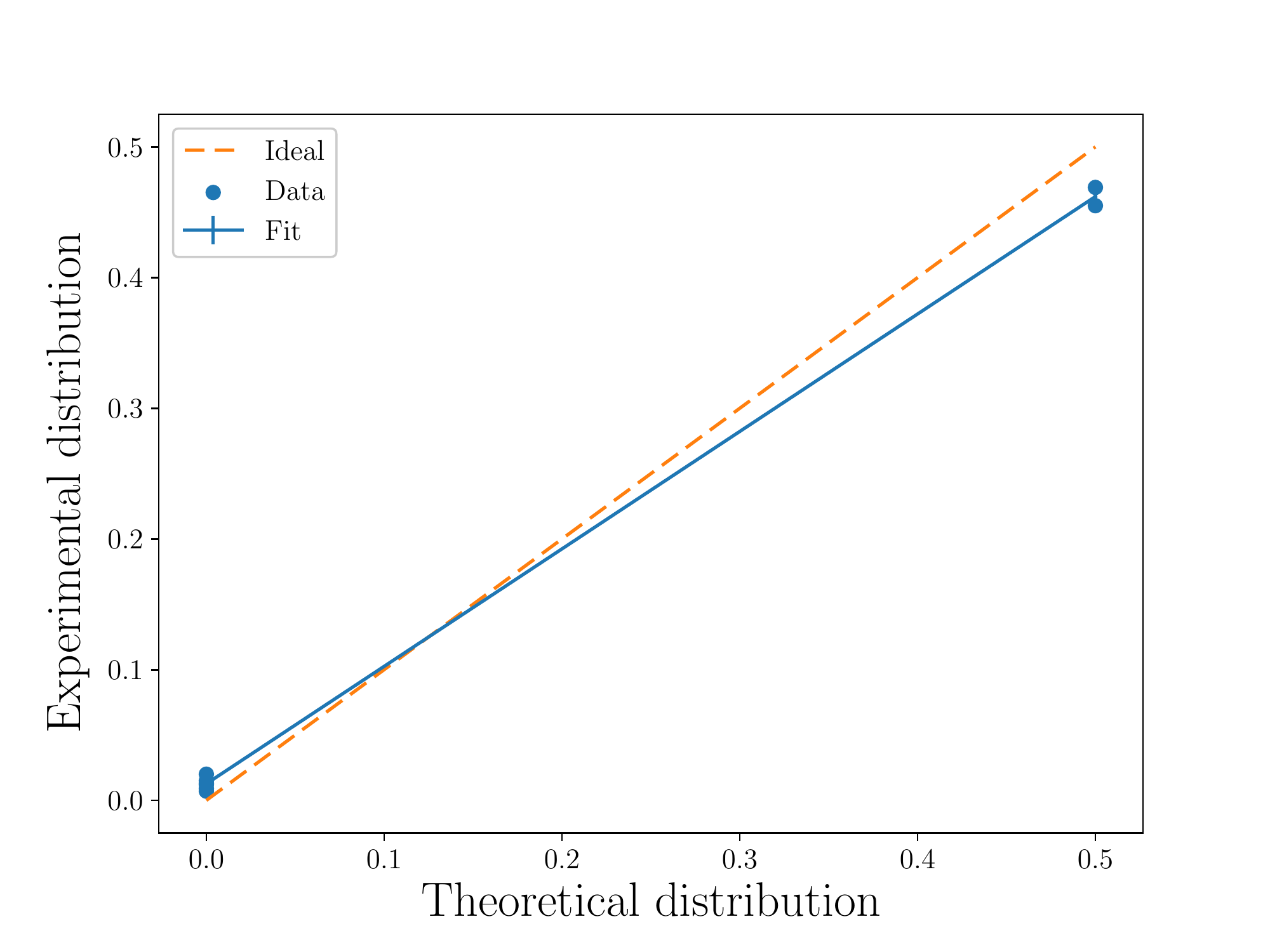}} \\
	\subfloat[]{\label{fig:sso_11_15}\includegraphics[width = 3.2in, height = 2.0 in]{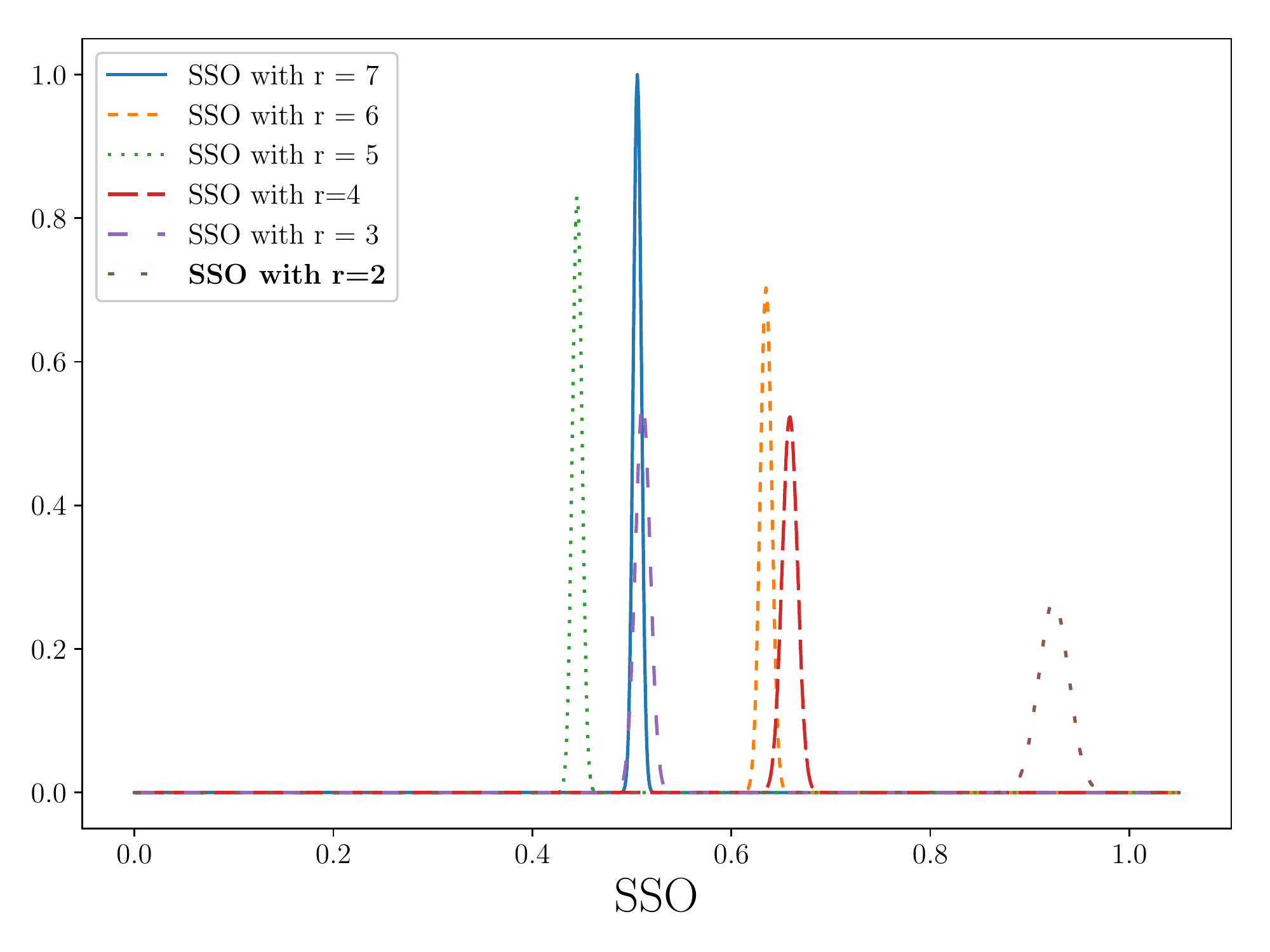}} 
	\caption{(a) Probability of finding a given phase for $N=15$ with base $a=11$, and (b) probability plot of the theoretical distribution and the experimental distribution. (c) SSO of the experimental data with the theoretical probability distributions corresponding to all possible periods.}
\label{histo_11_15}
\end{figure}

\begin{figure}[]
	\subfloat[]{\label{fig:hist_2mod21_2}\includegraphics[width = 3.2in, height = 2.0 in]{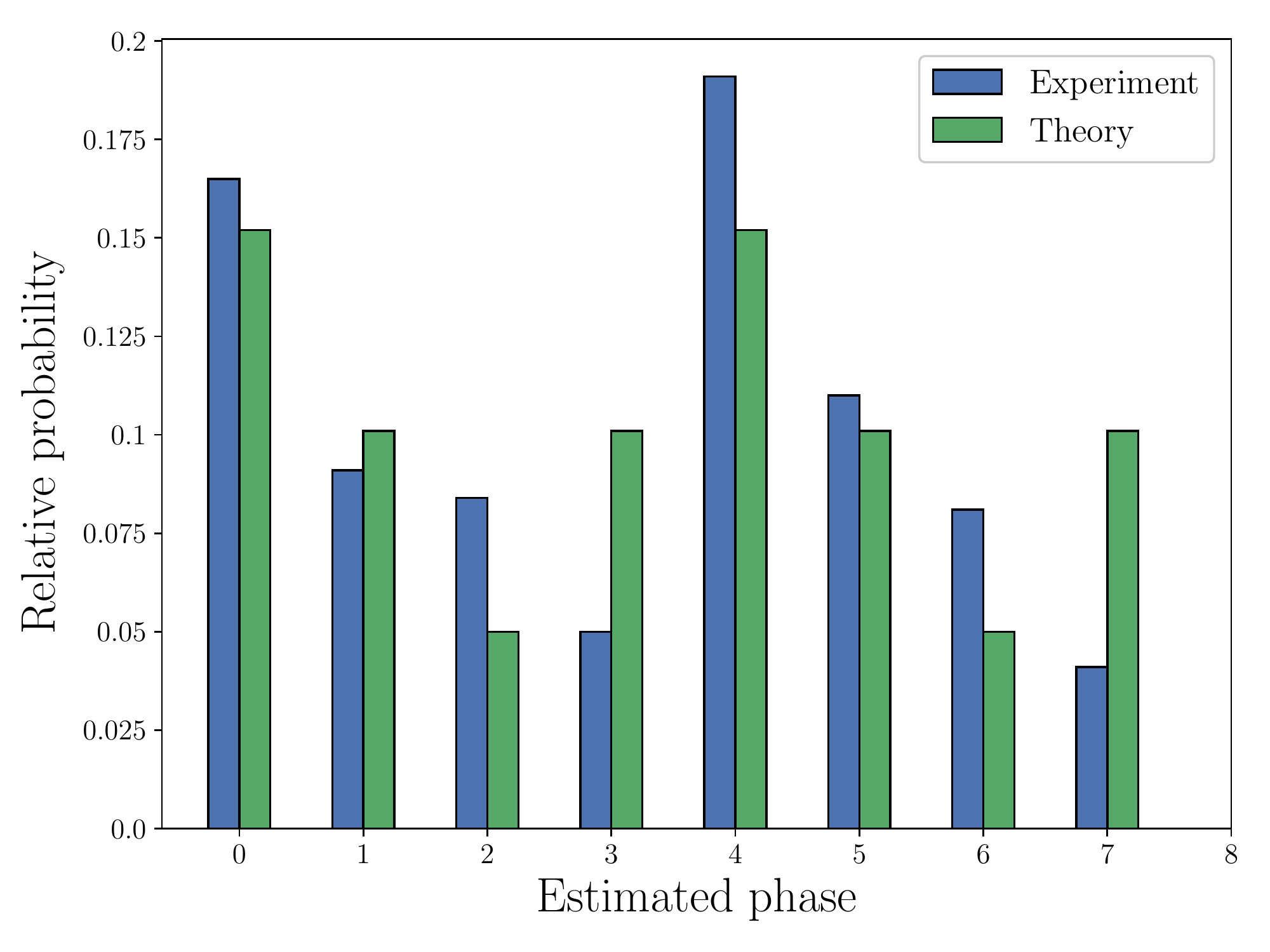}} \\ 
	\subfloat[]{\label{fig:pp_plot_2_21}\includegraphics[width = 3.6in, height = 2 in]{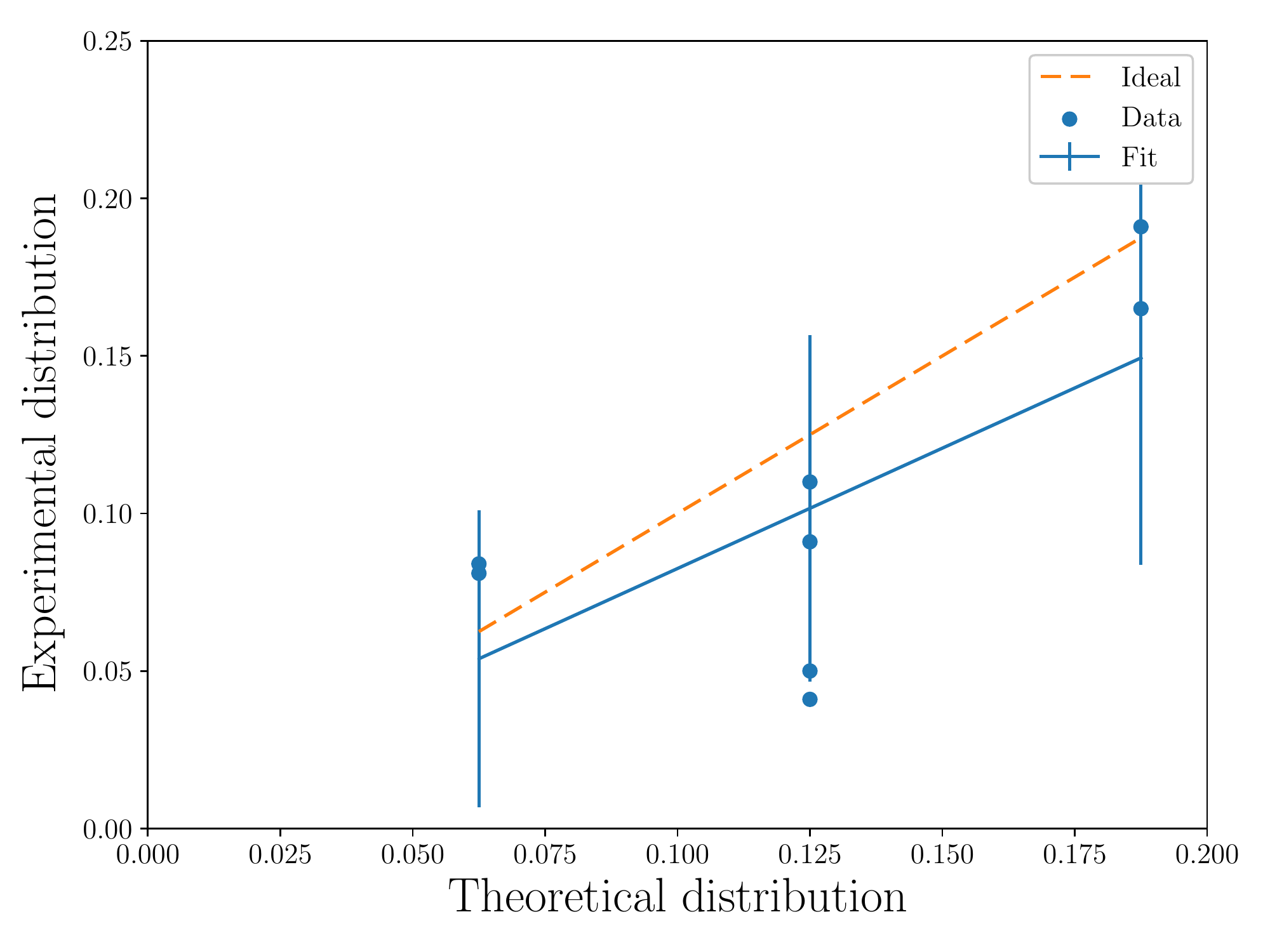}}\\
	\subfloat[]{\label{fig:sso_2_21}\includegraphics[width = 3.2in, height = 2.0 in]{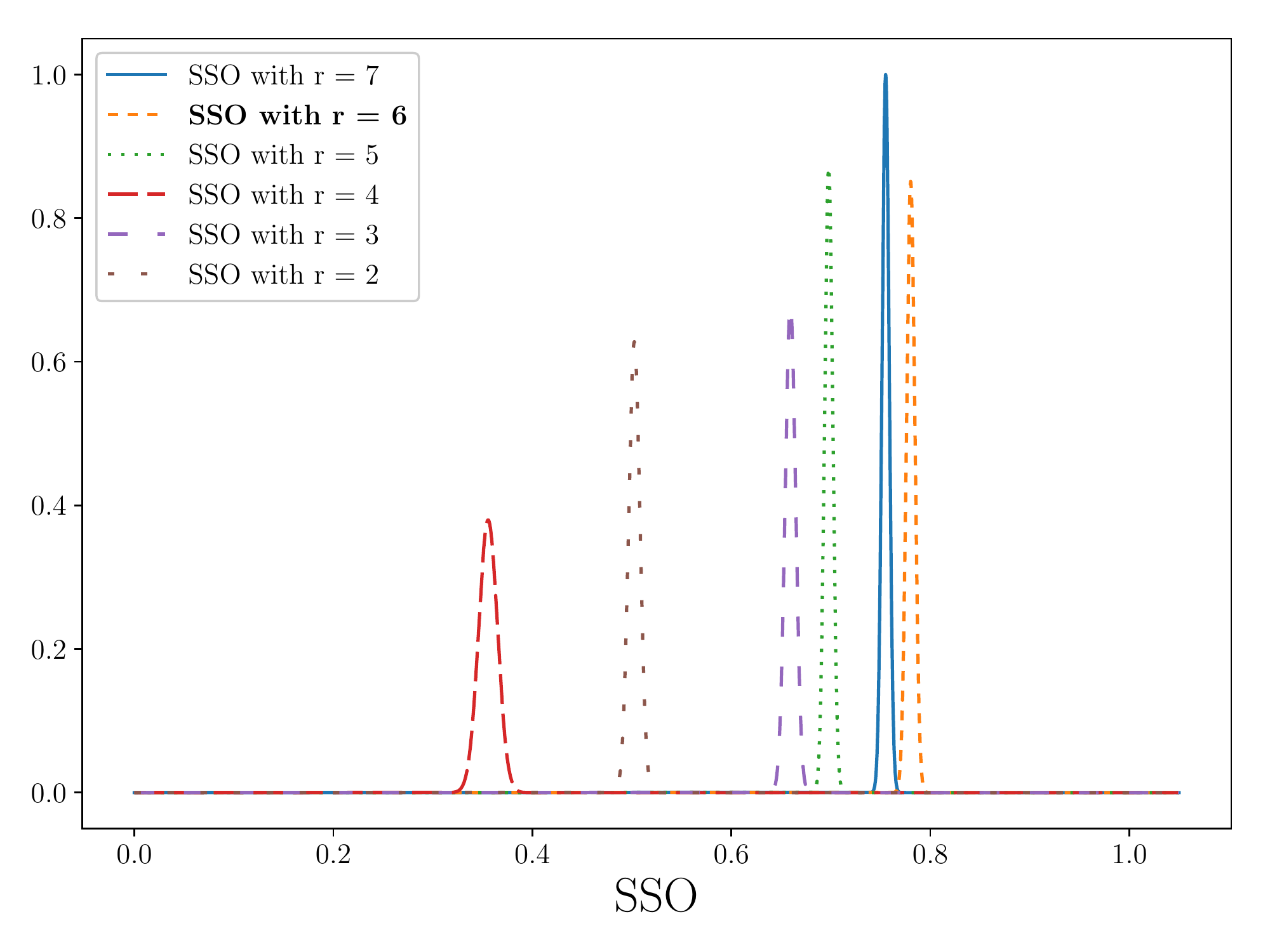}}  
	\caption{(a) Probability of finding a given phase for $N=21$ with base $a=2$, and (b) probability plot of the theoretical distribution and the experimental distribution. (c) SSO of the experimental data with the theoretical probability distributions corresponding to different periods.}
\label{fig:hist_2mod21}
\end{figure}

\begin{figure}[]
	\subfloat[]{\label{hist_4mod35}\includegraphics[width = 3.2in, height = 2.0 in]{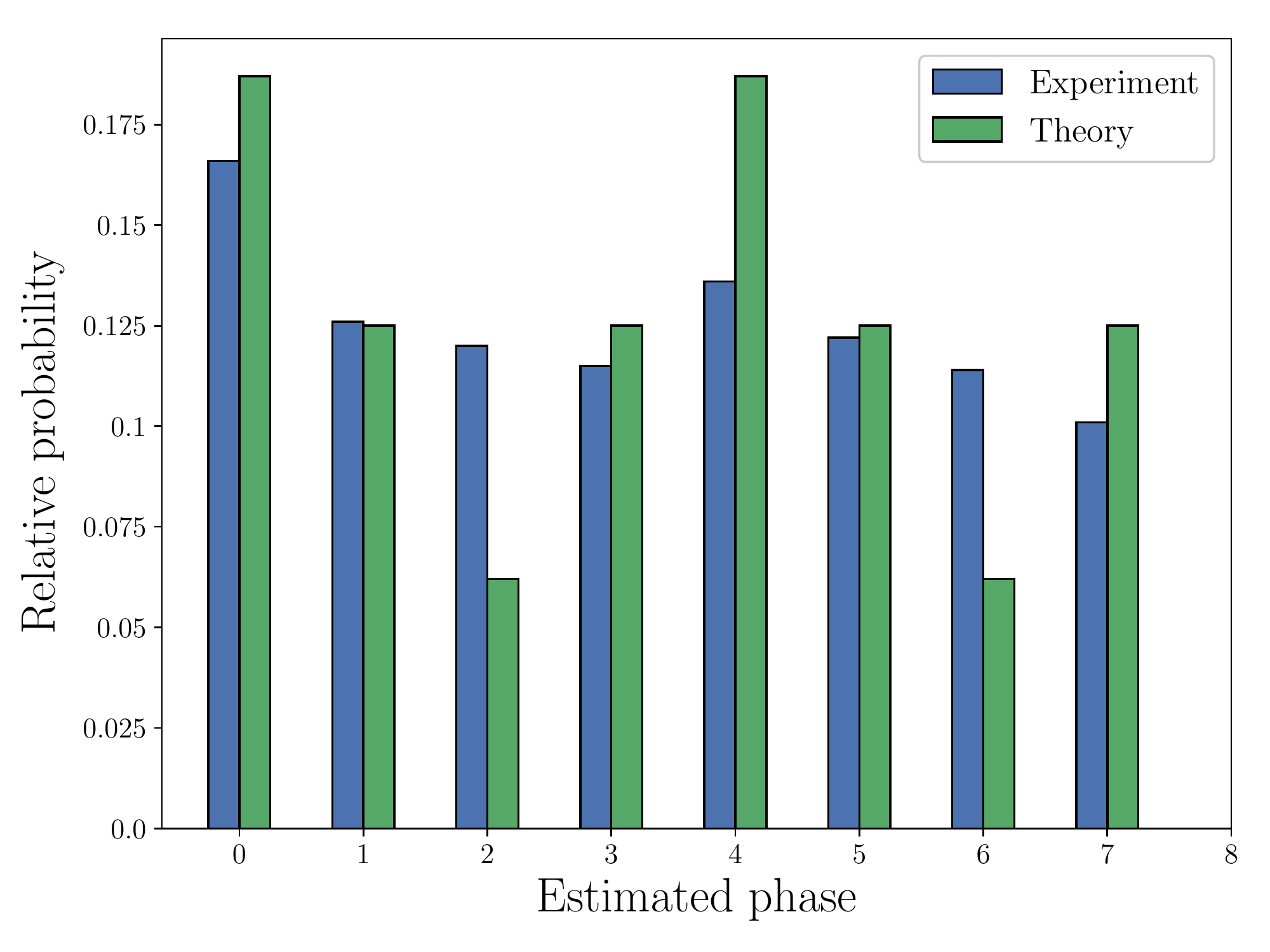}} \\ 
	\subfloat[]{\label{fig:pp_plot_4_35}\includegraphics[width = 3.2in, height = 2.0 in]{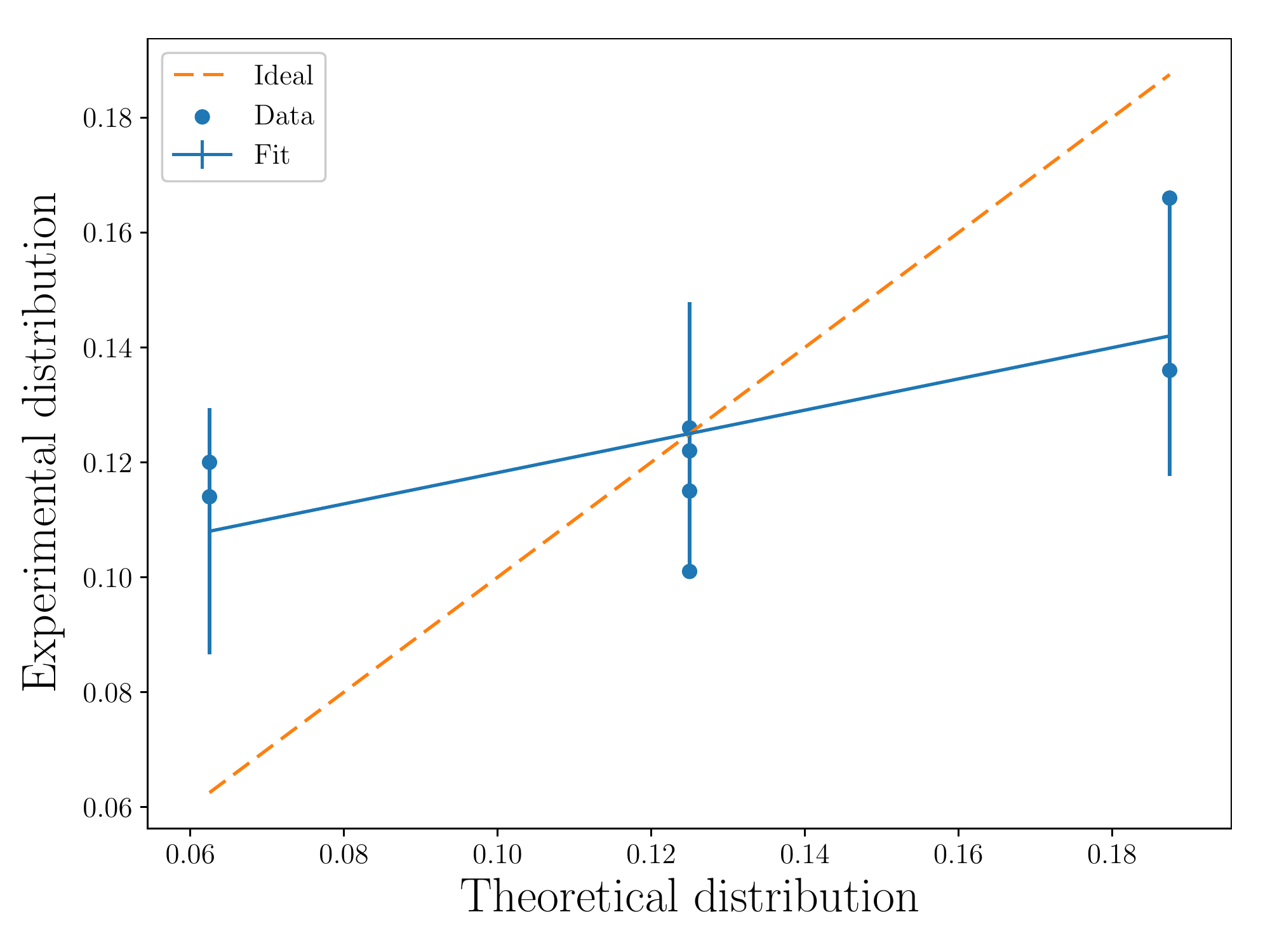}}\\
	\subfloat[]{\label{fig:sso_4_35}\includegraphics[width = 3.2in, height = 2.0 in]{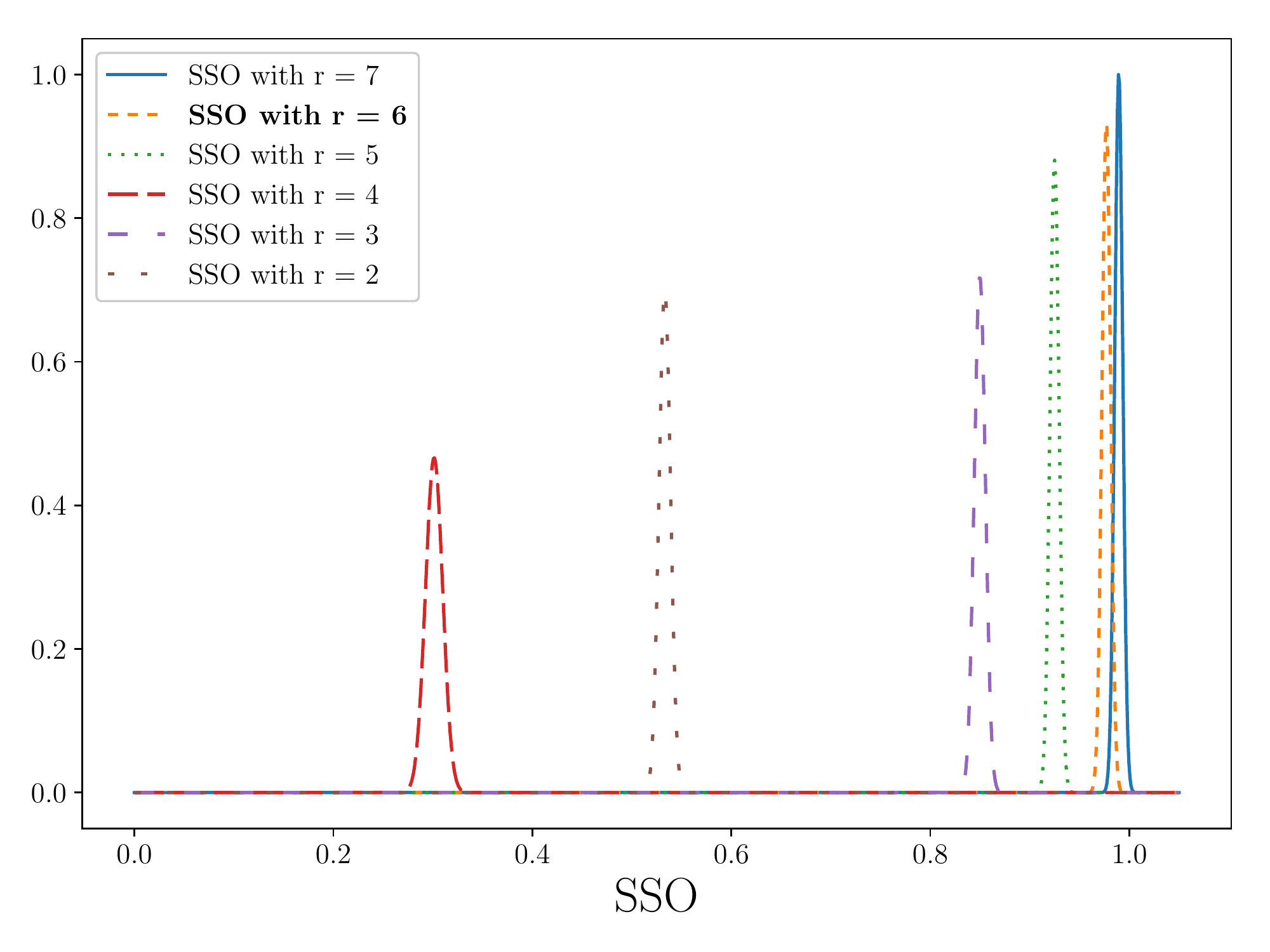}}  
	\caption{(a) Probability of finding a given phase for $N=35$ with base $a=4$, and (b) probability plot of the theoretical distribution and the experimental distribution. (c) Plot of the SSO between the experimental data and the all the possible theoretical distribution for the different values of $r$.}
\label{fig:hist_4mod35}
\end{figure}

The success of the experiment is evaluated in two different ways. We use probability plots to give a qualitative estimation of the correctness of the results, while the square of the statistical overlap (SSO) is used as a quantitative measure.
Probability plots \cite{kleiner} are a useful tool to visually compare two distributions. In a probability plot, one distribution is plotted against the other. If the two distributions are identical, the plot will show a straight line ($y=x$). The amount of deviation from the straight $y=x$ line is an indication of the difference between the two probability distributions plotted. For the case at hand, this means plotting on the $(x,y)$ plane a point for each value of the phase, where the value of the $x$ coordinate is given by the theoretical value of the probability distribution for that phase and the value of the $y$ coordinate is given by the corresponding experimental value found. The data are then fitted with a straight line for comparison with the ideal $y=x$ case. Error-bars on the fit are given as a range of y-values compatible with the error on the fit coming from both slope and offset of the fitted line at a fixed x-value. Thus, all straight lines contained within these error-bars are compatible with the experimental data within the estimated error for the fit. The probability plots between the experimental distribution and the expected theoretical one for each case are shown in Figs. \ref{fig:pp_plot_2_15}, \ref{fig:pp_plot_11_15}, \ref{fig:pp_plot_2_21} and \ref{fig:pp_plot_4_35}. In the case of $N=15$, the data in Figs. \ref{fig:pp_plot_2_15} and \ref{fig:pp_plot_11_15} are on a straight line very close to the $y=x$ line (tagged as "Ideal" on the plots). For the $N=21$ case, the data lie on a straight line parallel to the $y=x$ ideal line as can be seen from Fig. \ref{fig:pp_plot_2_21}. This means that there is an offset in the relative frequency of each phase in our experimental distribution. However, the overall shape coincides with the theoretical one, indicating that the difference in relative frequencies between phases is preserved. Finally, for $N=35$, the data in Fig. \ref{fig:pp_plot_4_35} lie on a straight line which is very far from the $y=x$ line, indicating an important deviation of the experimental results from the theoretically expected ones. In facts, looking at the histogram in Fig. \ref{hist_4mod35} shows that the experimental results were affected by noise, which tends to make all phases equally probable. 
In summary, for $N=15$ and $N=21$ the fit is close to the ideal line (within the error bars) but for $N=35$ it is not. Therefore, we believe that the probability plots provide a good qualitative measure of the similarity between probability distributions, as they correctly describe the similarity which is apparent by the comparison of the histograms of the distributions.

Next we give a quantitative measure of the correctness of the results. In particular, we want to answer the question: given the experimental data obtained, what is the likelihood that this data comes from a given probability distribution?
The answer to this question will reveal two aspects of our experiment. First, it will allow us to assign a period to the results without the need for the continued fraction algorithm. Second, it will give us a measure of the error we make in the assignment. Our method of assigning the period to the experimental data relies on the following observation: the probability of obtaining a certain phase $s$ is

\begin{equation}
\text{P}(s) = \frac{1}{Q} \lvert e^{\frac{2\pi i x_0}{Q}} \rvert^2 \sum_{s=0}^{Q-1}  \lvert e^{\frac{2\pi i s d r}{Q}} \rvert^2,
\end{equation}

\noindent
independently of the base $a$ chosen. As a function of the estimated phase $s$, the probability distribution $\text{P}(s)$ is completely characterized by the values of the parameters $r$ and $Q$, the period and the number of bits $\text{log}_2 Q$ used to encode the value of the period, respectively. Therefore, independently of the base chosen, there correspond a fixed probability distribution to each value of $r$ and $Q$. To determine the period to assign to the experimental data, we compare the probability distribution $\text{P}_{\text{exp}}(s)$ obtained experimentally, with all the possible probability distributions $\text{P}^{r}_{\text{th}}(s)$ given by values of $r$ from $2$ to $Q-1$, for fixed number of bits $\text{log}_2 Q$ encoding the period. The period of the theoretical distribution which is most similar to the experimental data is then assigned to the experiment. 

Following \cite{monz}, we use the square of the statistical overlap (SSO) introduced in \cite{chiaverini} as a measure of similarity between probability distributions. The SSO is defined as 

\begin{equation}
\text{SSO} = \left(\displaystyle{\sum_{j=0}^7} m_j^{1/2} e_j^{1/2}\right)^2,
\end{equation}

\noindent
where $m_j$ and $e_j$ are the measured and expected output-state probabilities of state $\lvert j \rangle$, respectively.  

One can calculate the error on the SSO from the Poissonian counting error of the data, assuming Gaussian propagation of errors

\begin{equation}
\Delta\text{SSO} = \sqrt{\displaystyle{\sum_{j=0}^7} \frac{\partial}{\partial m_j}(m_j^{1/2} e_j^{1/2})^2\Delta m_j^2}.
\end{equation} 

\noindent
For each base used in the experiments, we calculate the SSO of $\text{P}_{\text{exp}}(s)$ with all possible $\text{P}^r_{\text{th}}(s)$. To better visualize which $\text{P}^{r}_{\text{th}}(s)$ most resembles the data, we plot unit area normalized Gaussian distributions with the SSO as the mean and $\Delta \text{SSO}$ as the standard deviation. The Gaussian whose value of the mean is closest to 1 comes from the $\text{P}^{r}_{\text{th}}(s)$ most similar to $\text{P}_{\text{exp}}(s)$. Therefore, we assign period $r$ to our data. While the spread of each Gaussian gives an indication of the error in the calculation of the SSO. To quantitatively determine the error in the assignment of the period, we calculate the area of overlap between the Gaussian distribution with the highest SSO and the second closest one. This is done through the overlap coefficient \cite{inman} between the normal distributions. The OVL is defined as

\begin{equation}
\text{OVL}\left[ f(x_1), f(x_2) \right]= \sum_{x} min{\left(f(x_1), f(x_2)\right)},
\end{equation}

\noindent
where $f(x_1)$ is the normal distribution with the highest SSO and $f(x_2)$ is the normal distribution with second highest SSO. The OVL tells us what is the probability that the assignment is done incorrectly i.e. the highest SSO for our experimental data comes from a different theoretical probability distribution than the assigned one. Thus, we quantify the error on our assignment as $\epsilon_{ij} \equiv \text{OVL}\left[ f(x_i), f(x_j) \right]$ where $i$ denotes the period of the distribution with the highest SSO and $j$ the period of the distribution with the second highest SSO.

The results of the comparison for all experiments are presented in Figs. \ref{fig:sso_2_15}, \ref{fig:sso_11_15}, \ref{fig:sso_2_21}, and \ref{fig:sso_4_35}. Figs. \ref{fig:sso_2_15} and \ref{fig:sso_11_15} show the SSO of the experimental distributions and their deviations obtained for $N=15$, $a=2$ and $a=11$, respectively. For $a=2$, the highest SSO is $0.97$ for the theoretical distribution corresponding to the period $r=4$. Thus, we assign the period $r=4$ to the experimental distribution obtained. The error we make in assigning the period $r=4$ instead of period $r=7$, which is the closest match, is $\epsilon_{47} = 3.8 \times 10^{-134}$. For $a=11$, the highest SSO is $0.92$ which correspond to $r=2$. The error in the assignment of $r=2$ with respect to $r=4$, which has the second highest SSO, is $\epsilon_{24} = 4.1 \times 10^{-31}$.
The results obtained for $N=21$ with $a=2$ are shown in Fig. \ref{fig:sso_2_21}. Here, it is more difficult to determine the period with certainty. The highest SSO is $0.78$, which corresponds to the theoretical distribution with $r=6$. The error in assigning $r=6$ to the experimental data is $\epsilon_{67} = 1.2  \times 10^{-3}$. Therefore, there is a $\sim 0.1\%$ chance that we assigned the period incorrectly and the true period was $r=7$ instead.
For the case of $N=35$ and $a=4$, the results presented in Fig. \ref{fig:sso_4_35} show that the highest SSO between the experimental data and the theoretical distribution corresponding to all possible periods is $0.99$ for $r=7$, although this is not the expected period. There is another close match with an SSO of $0.98$ for $r=6$, which is the correct one. The error in assigning period $r=7$ to the experimental data instead of $r=6$ is $\epsilon_{76} = 0.14$. Thus, in this case it is quite difficult to discern the correct period.

\section{Conclusions}
\label{conclusions}
Although the results are obtained with a compiled and simplified version of Shor's factoring algorithm, our purpose is to show a way to proceed with the implementation of generic algorithms on the approximate quantum computers available now. In practice, the non-negligible noise and the lack of key functions of the device force us to rethink how to design algorithms that can work on these machines. As it is evident from this work, one needs to supplement the deficiencies of the hardware with a more detailed theoretical analysis and classical processing. By doing so, one can reduce the length of the circuit needed to implement the algorithm, mitigating the effects of noise, and overcoming the lack of particular functions assumed for a general-purpose quantum computer. We emphasize that the simplification by inspection done here was possible only due to the small size of the circuit. Larger circuits would require a more sophisticated optimization.
We used different methods to evaluate the success of the experiment. The first one is the probability plot, which gives a qualitative measure of the similarity between the distribution of the experimental data and the expected theoretical distributions. The second one is the SSO, which gives a quantitative measure of the similarity between probability distributions. By using the SSO, we introduced a new way to assign a certain period to the probability distribution obtained from the experimental data. In this way, we avoid using the continued fraction algorithm, which fails when the number of bits used to encode the value of the period is particularly low, as in our situation. To correctly quantify the error which can be made in this assignment, the OVL between different candidates for the period is calculated. Overall, the experimental results obtained from running the algorithm on the \textit{ibmqx5} device are in agreement with the theoretical expectation values. Excellent agreement is found for $N=15$, while deviations from the theoretical results become more noticeable for $N=21$. Eventually, the algorithm fails to factor $N=35$. This is due to the cumulative errors coming from the increasing number of two-qubits gates necessary to implement the more complex MEF needed for this case.

\acknowledgments
We acknowledge use of the IBM Q for this work. The views expressed are those of the authors and do not reflect the official policy or position of IBM or the IBM Q team.
The authors are grateful to N. T. Bronn and R. Ya. Kezerashvili for the valuable and stimulating discussions.

\appendix
\section{circuits for the MEF}
\label{app_A}

Here we present the procedure used to implement the MEF in the experiments for factoring $N=15,21$ and $35$. These were specifically designed to reduce the number of gates to the minimum and mitigate the effects of noise. To make the approach scalable, one would need an automatic way to generate the modular exponentiation circuits as proposed in Ref. \cite{monz}. 

\begin{figure}[H]
	\subfloat[]{
		\label{circuit_15_a_4}
		\includegraphics[height = 1.2 in]{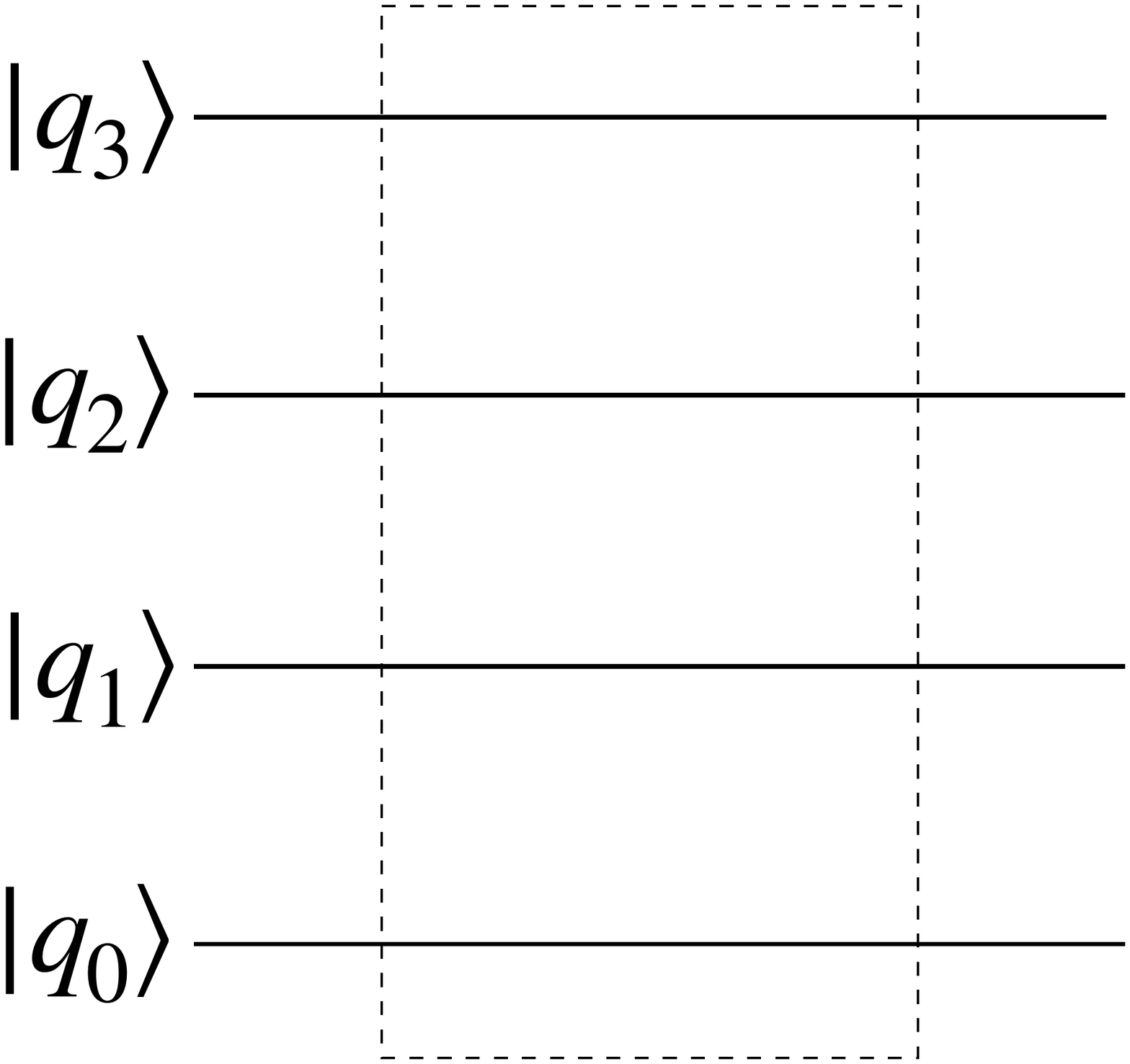}
	}
	\subfloat[]{
		\label{circuit_15_a_2}
		\includegraphics[height = 1.2 in]{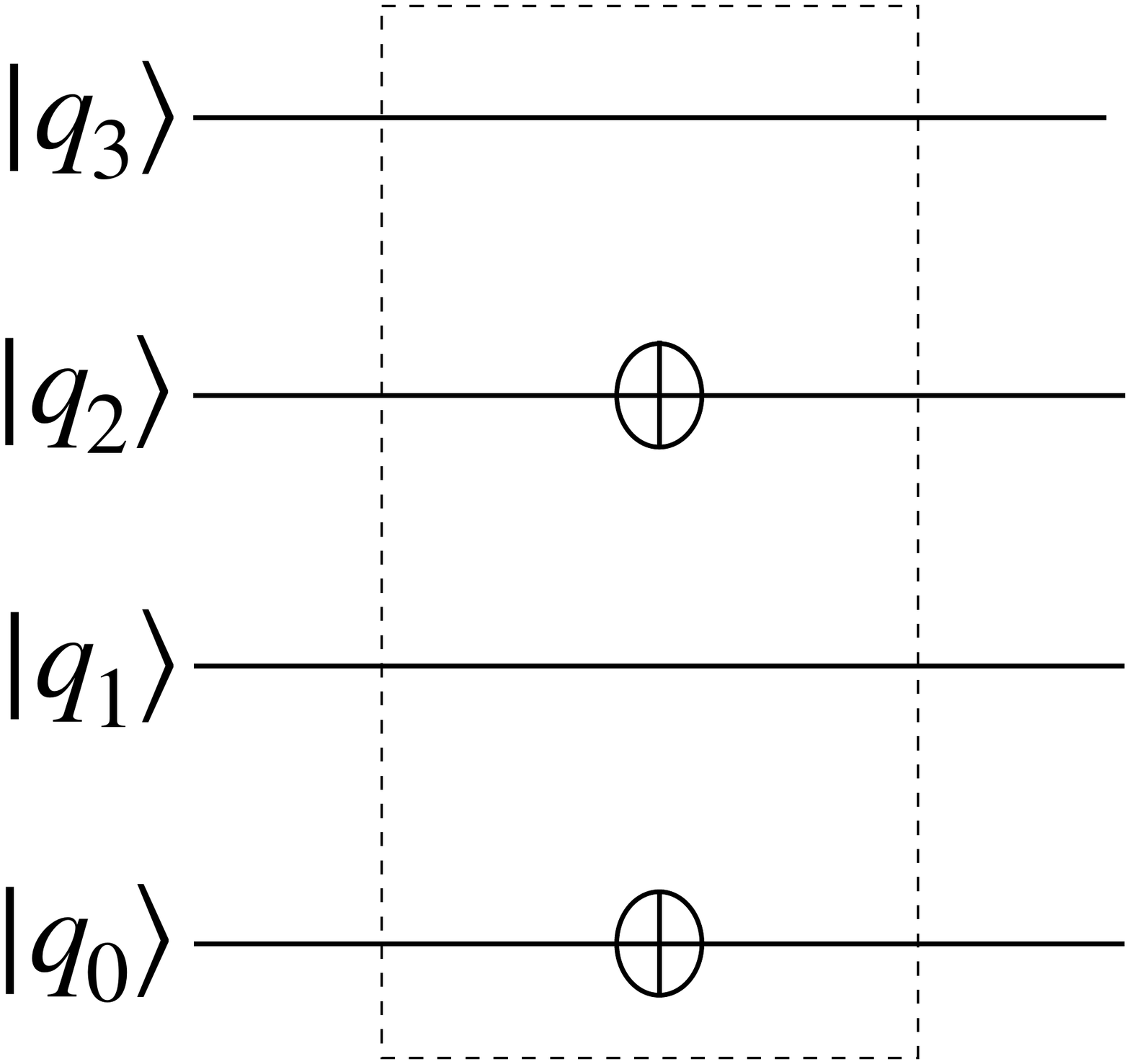}
	}
	\\	
	\subfloat[]{
		\label{circuit_15_2_1}
		\includegraphics[height = 1.2 in]{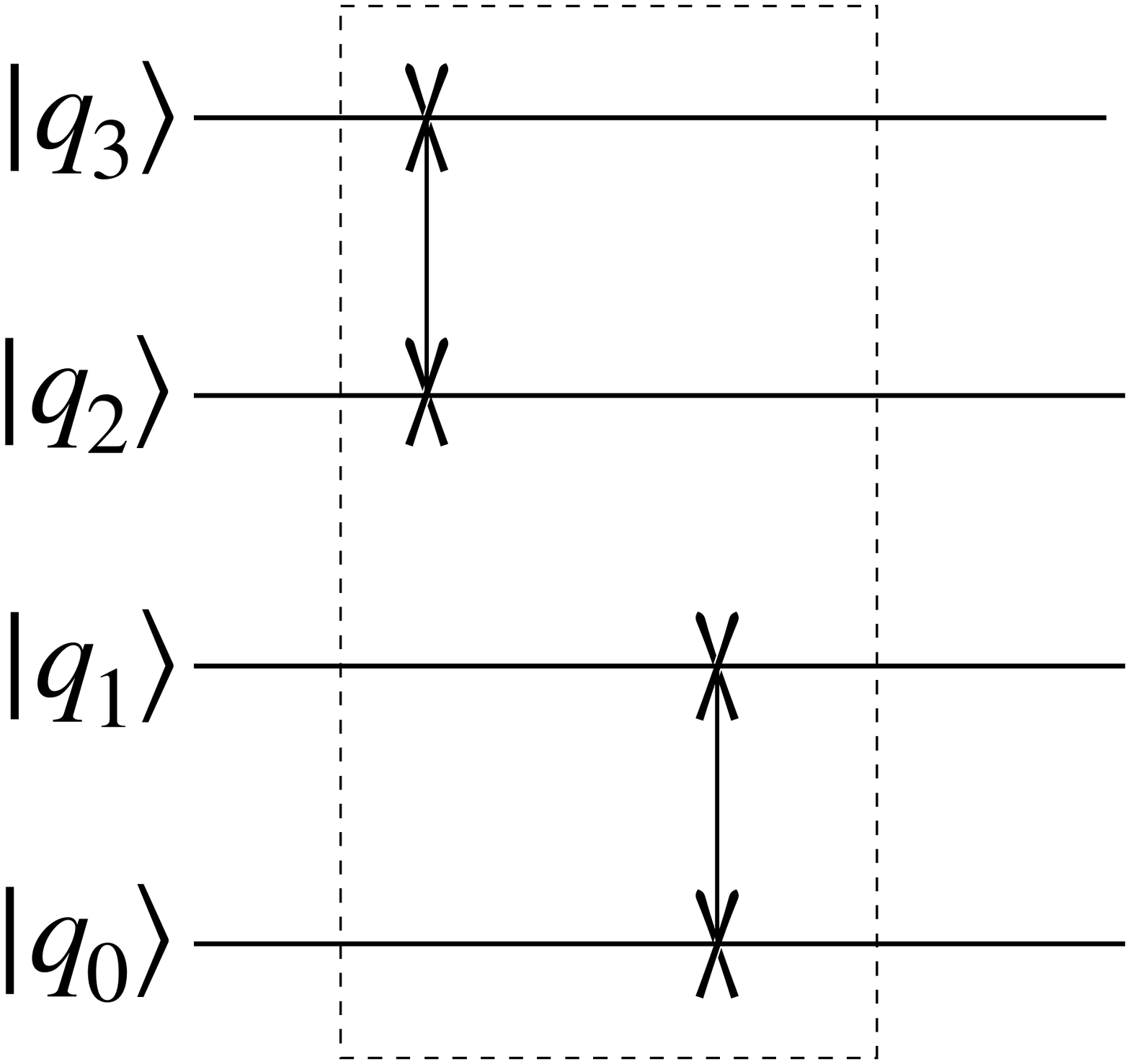}
	}
	\subfloat[]{
		\label{circuit_15_11_1}
		\includegraphics[height = 1.2 in]{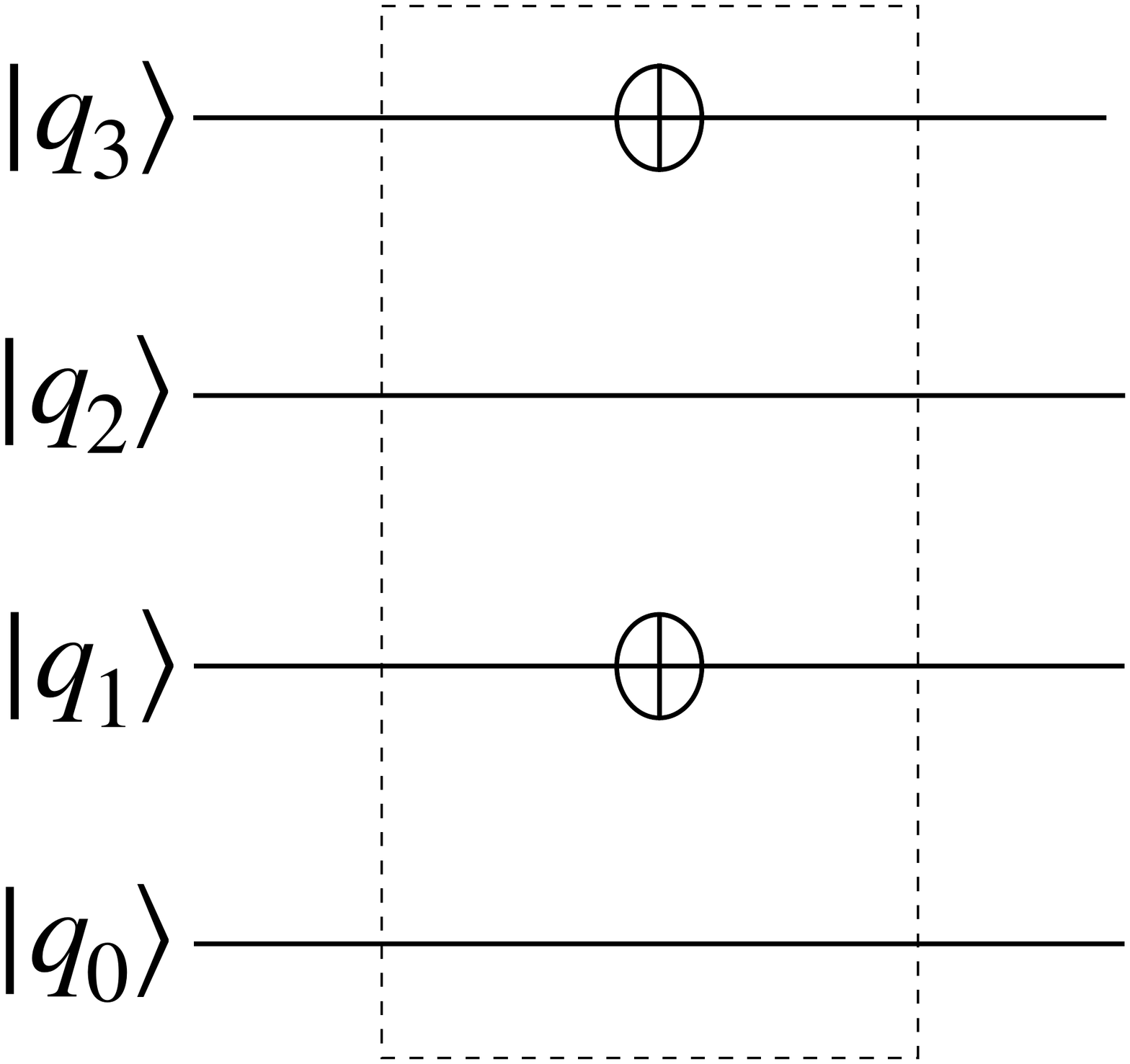}
	}		
	\caption{Modular exponentiation circuits for $N=15$. (a) $a^4 \,  \text{mod} \, 15$ for any $a$ and $a^2 \,  \text{mod} \, 15$ for $a=11$, (b) $a^2 \,  \text{mod} \, 15$ for $a = \left\{2,7,8,13\right\}$, (c) $2^1 \,  \text{mod} \,  15$ for $a = 2$, and (d) $11^1 \,  \text{mod} \, 15$ for $a = 11$ .}
	\label{mod_exp_15}	
\end{figure}

\noindent
The circuits used for the MEF in the experiment for factoring $N=15$ are shown in Fig. \ref{mod_exp_15}. The MEF in the first circuit of Fig. \ref{fig:full_circuit_broken} shown in Fig. \ref{circuit_15_a_4} is the identity operation for any base $a$, making it a deterministic step. The output of the first circuit, is then fed into the second one. As shown in Ref. \cite{monz}, the MEF here reduces to a very simple circuit depending on the base $a$ selected for factoring. If the base $a$ is any one of the elements of the set $\left\{4,11,14\right\}$, the modular exponentiation function is again the identity shown in Fig. \ref{circuit_15_a_4} and this step turns again into a deterministic step. If the base is one of the elements of the set $\left\{2,7,8,13\right\}$, the MEF has the same simple circuit for any of these $a$, which can be seen from Fig. \ref{circuit_15_a_2}. The MEF for the two bases $a=2$ and $11$ for the third circuit are given in Figs. \ref{circuit_15_2_1} and \ref{circuit_15_11_1}, respectively.

\begin{figure}[H]
	\subfloat[]{
		\label{circuit_21_2_4}
		\includegraphics[height = 1.2 in]{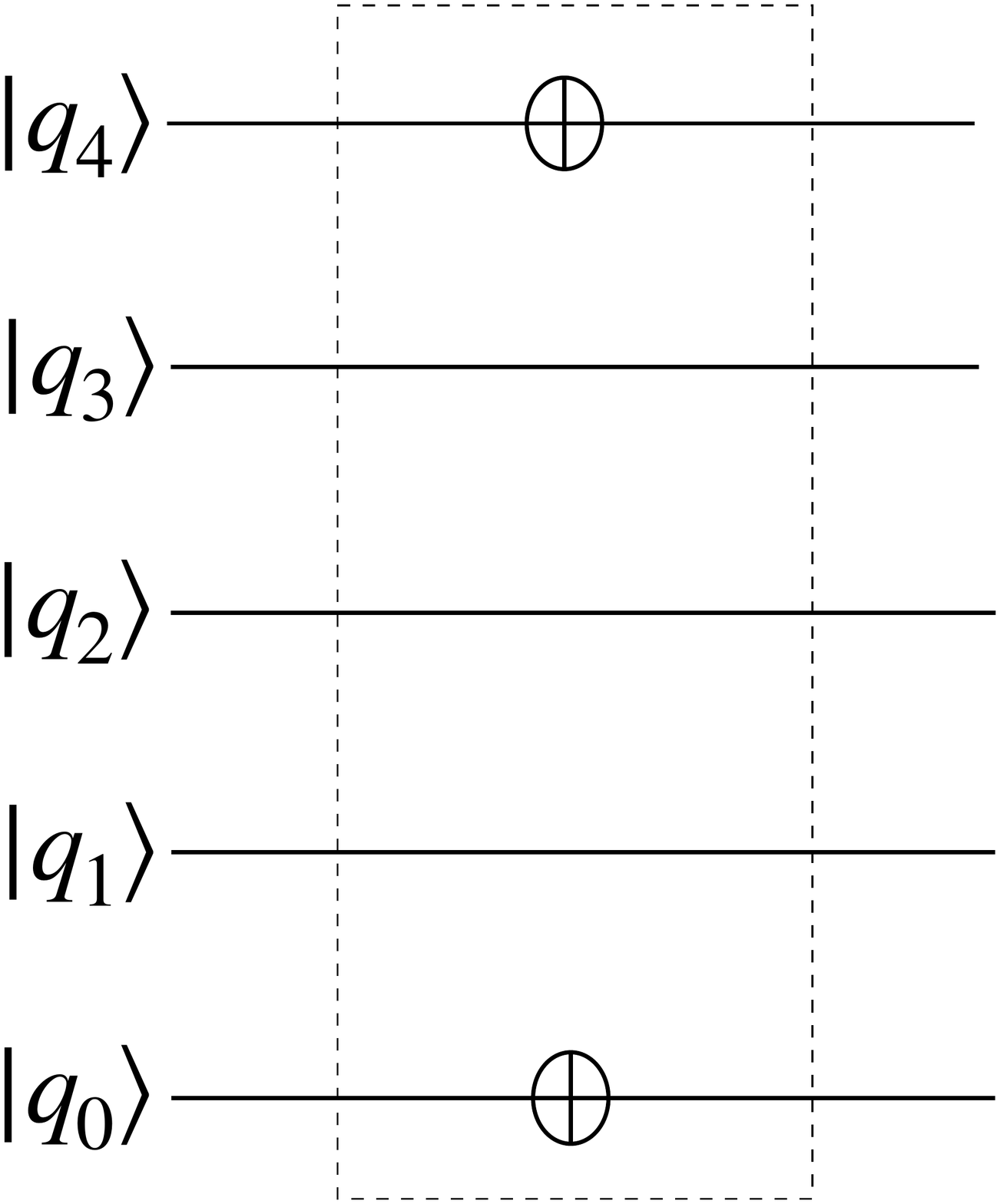}
	}
	\subfloat[]{
		\label{circuit_21_2_2}
		\includegraphics[height = 1.2 in]{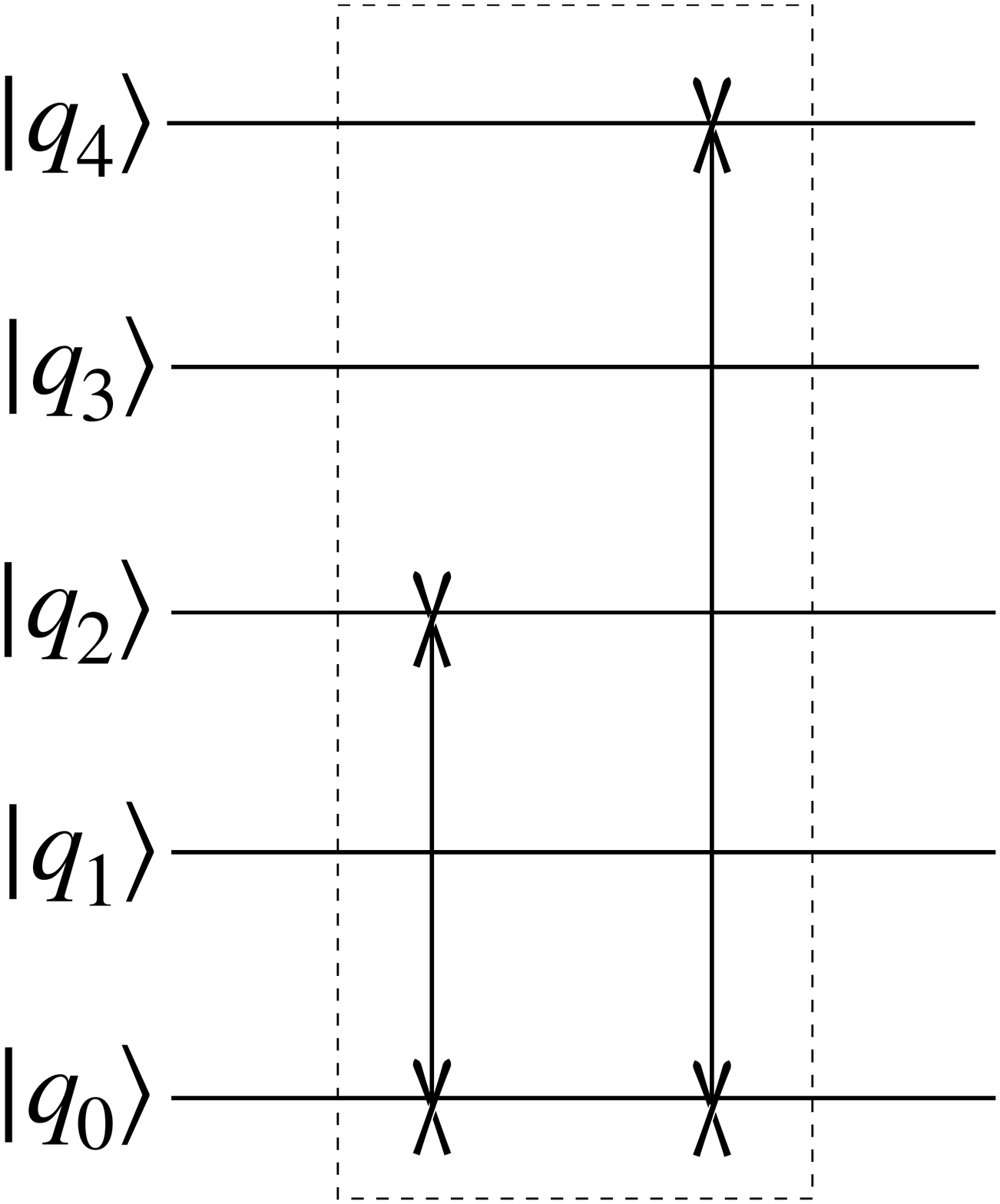}
	}
	\\	
	\subfloat[]{
		\label{circuit_21_2_1_00}
		\includegraphics[height = 1.2 in]{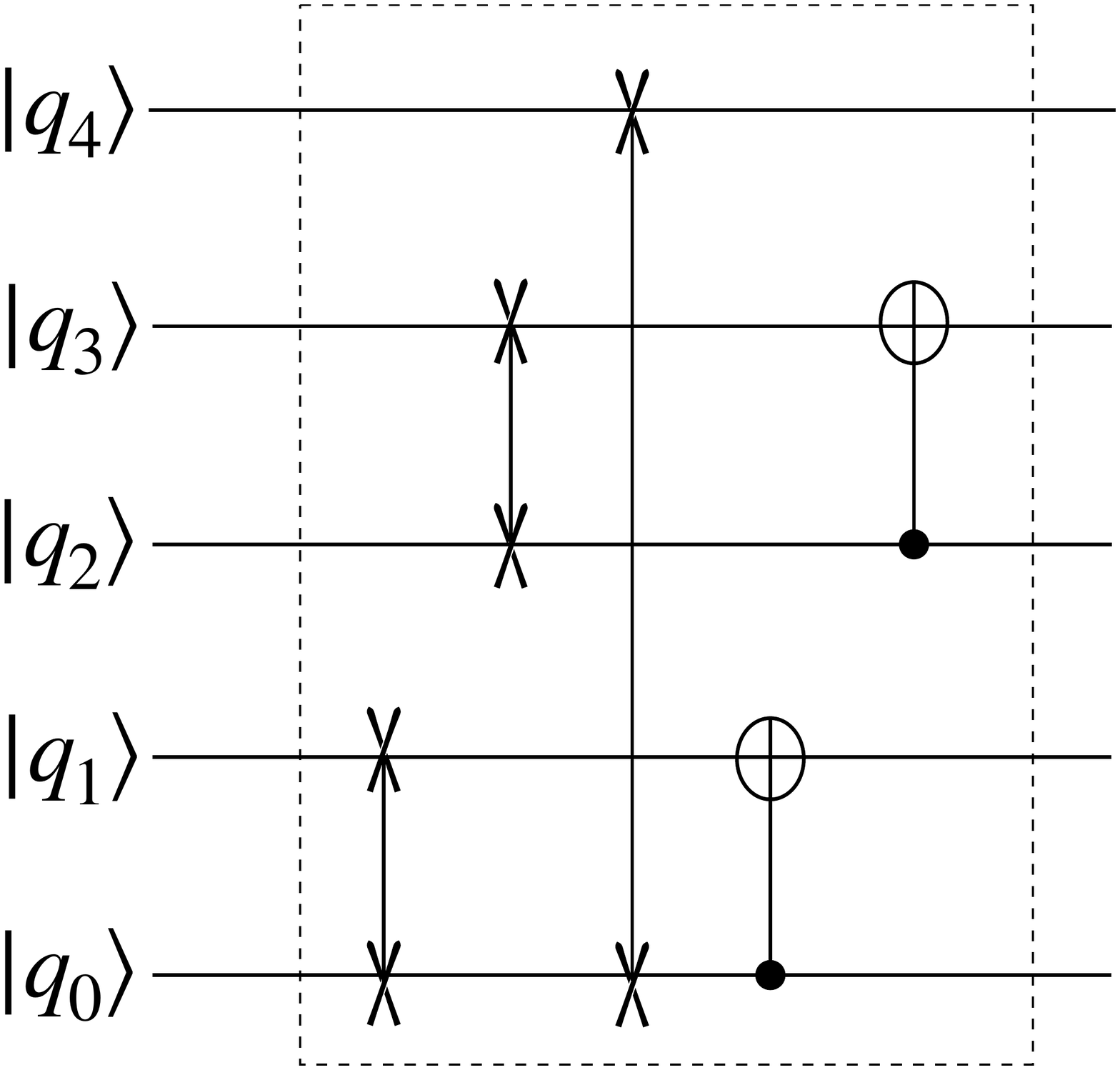}
	}
	\subfloat[]{
		\label{circuit_21_2_1_10}
		\includegraphics[height = 1.2 in]{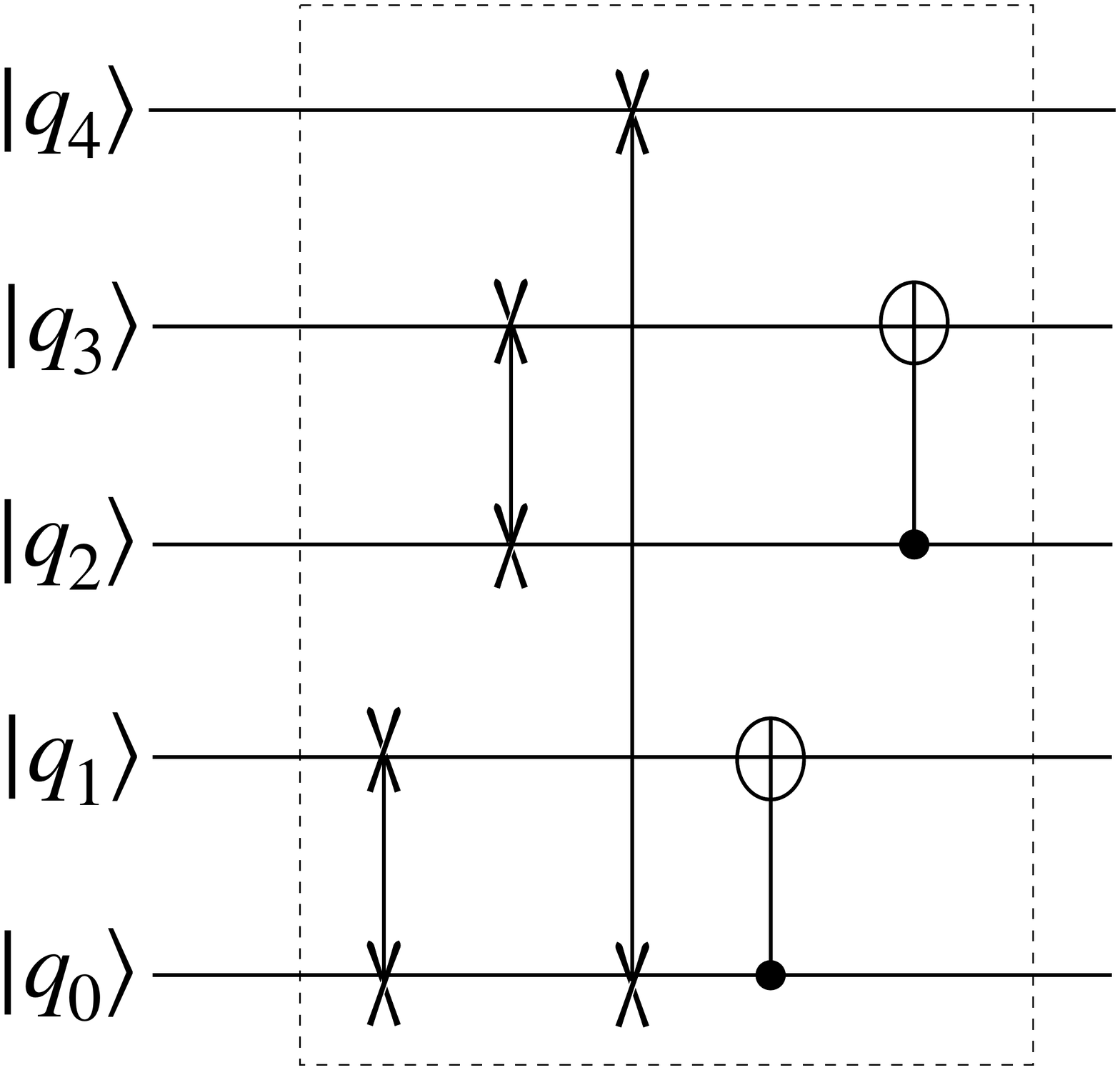}
	}
	\\
	\subfloat[]{
		\label{circuit_21_2_1_01}
		\includegraphics[height = 1.2 in]{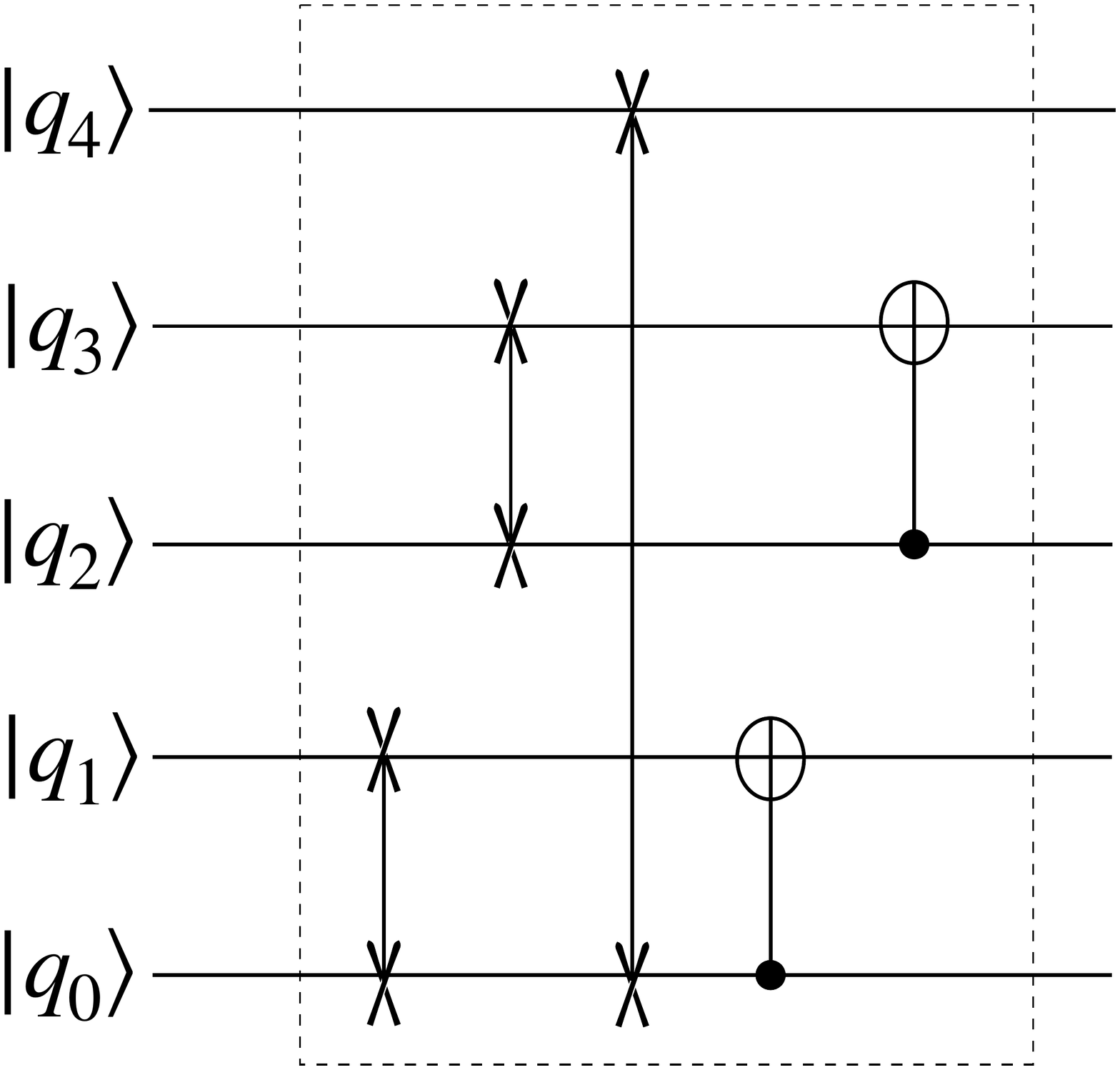}
	}
	\subfloat[]{
		\label{circuit_21_2_1_11}
		\includegraphics[height = 1.2 in]{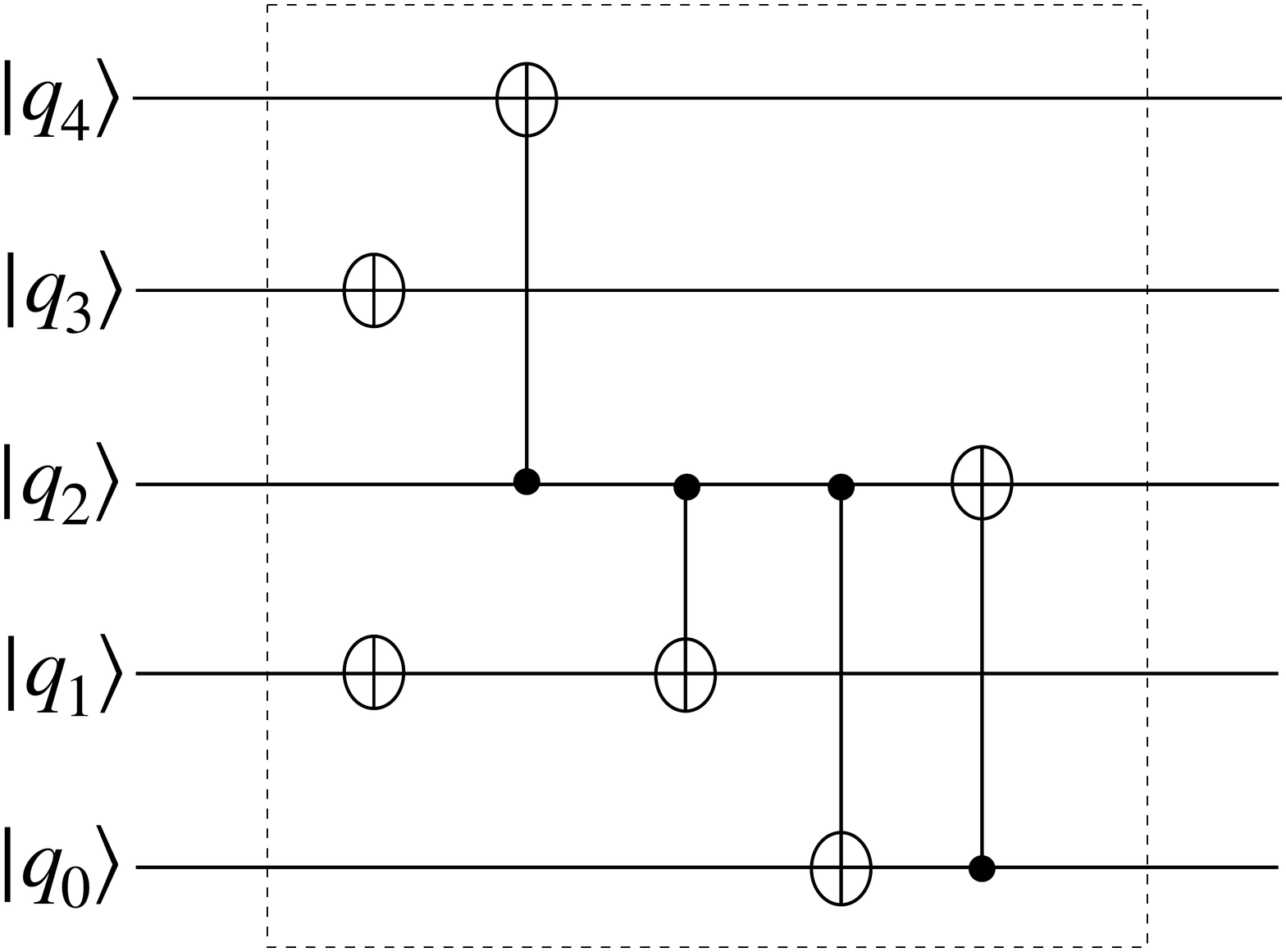}
	}		
	\caption{Modular exponentiation circuits for $N=21$ with base $a=2$. (a) $2^4 \,  \text{mod} \, 21$, (b) $2^2 \,  \text{mod} \, 21$. Depending on the results of the measurement of the period register in the previous circuits we have: (c)  $2^1 \,  \text{mod} \,  21$ for $\text{bit}^{(0)} = 0$ and $\text{bit}^{(1)} = 0$; (d) $2^1 \,  \text{mod} \, 21$ for $\text{bit}^{(0)} = 1$ and $\text{bit}^{(1)} = 0$, (e) $2^1 \,  \text{mod} \, 21$ for $\text{bit}^{(0)} = 0$ and $\text{bit}^{(1)} = 1$, and (f) $2^1 \,  \text{mod} \, 21$ for $\text{bit}^{(0)} = 1$ and $\text{bit}^{(1)} = 1$.}
	\label{mod_exp_21}	
\end{figure}

The circuits of the MEF used in the experiment of factoring $N=21$ are presented in Fig. \ref{mod_exp_21}. The experiment was conducted only with the base $a=2$, therefore all circuits have been designed only for this base. The MEF for the first circuit is shown in Fig. \ref{circuit_21_2_4}. For the second circuit, the MEF in Fig. \ref{circuit_21_2_2} was used. In the third circuit, depending on the values of the bits of the period register measured in the previous stages, different states are prepared as input. For this reason, different modular exponentiation circuits are designed according to the results of the measurements of the period register. The various possibilities are shown in Figs. \ref{circuit_21_2_1_00}, \ref{circuit_21_2_1_10}, \ref{circuit_21_2_1_01} and \ref{circuit_21_2_1_11} corresponding to the four possible outcomes $00$, $01$, $10$ and $11$, respectively.

The MEFs implemented in the experiment of factoring $N=35$ are depicted in Fig. \ref{mod_exp_35}. The circuits are designed for the algorithm with base $a=4$. The MEF for first, second and third circuits are shown in Figs. \ref{circuit_35_4_4}, \ref{circuit_35_4_2}, and \ref{circuit_35_4_1}, respectively. In this case, one circuits which works for any input was designed for the MEF at each stage.

\begin{figure}[H]
	\subfloat[]{
		\label{circuit_35_4_4}
		\includegraphics[height = 1.2in]{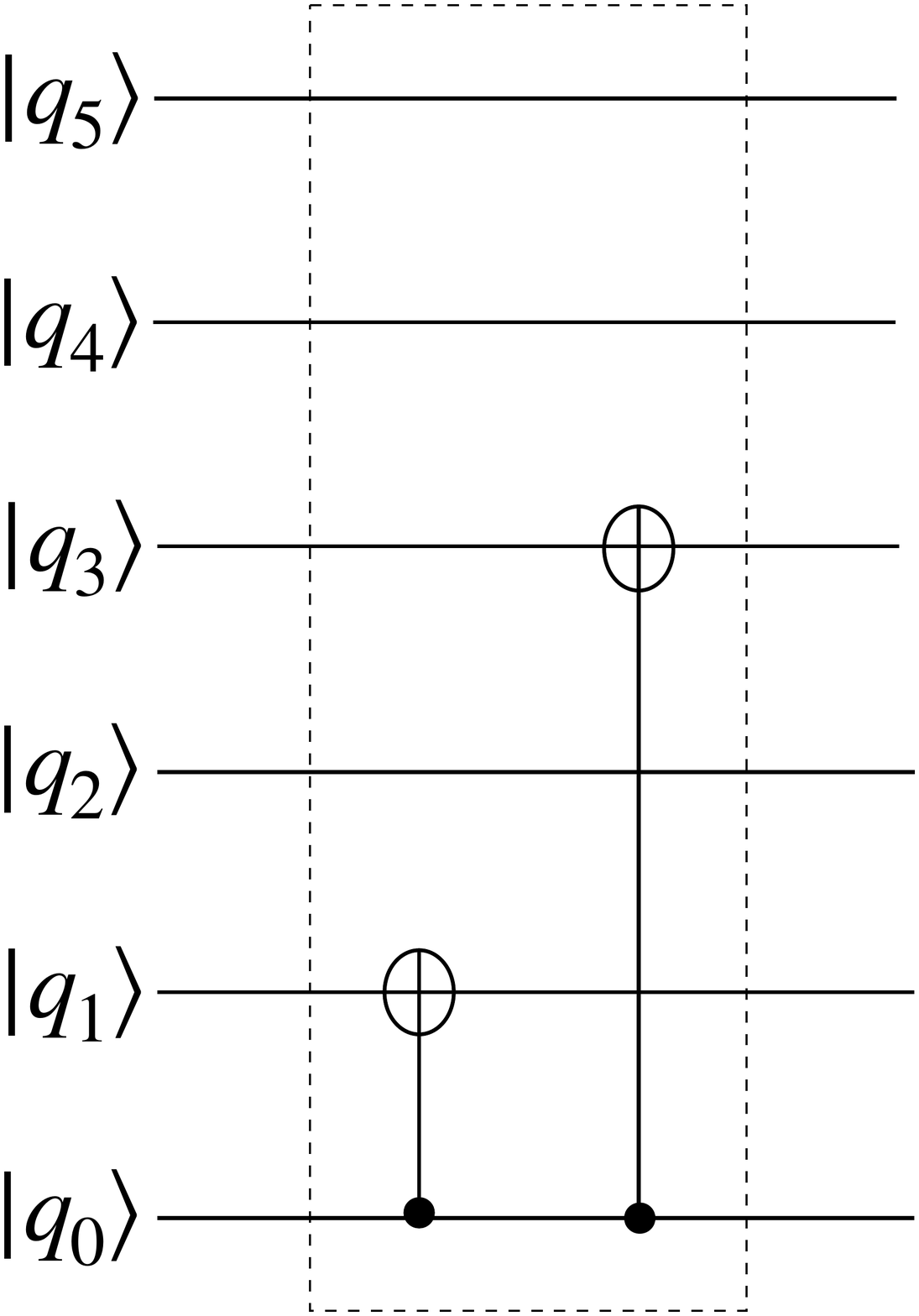}
		}
	\subfloat[]{
		\label{circuit_35_4_2}
		\includegraphics[height = 1.2 in]{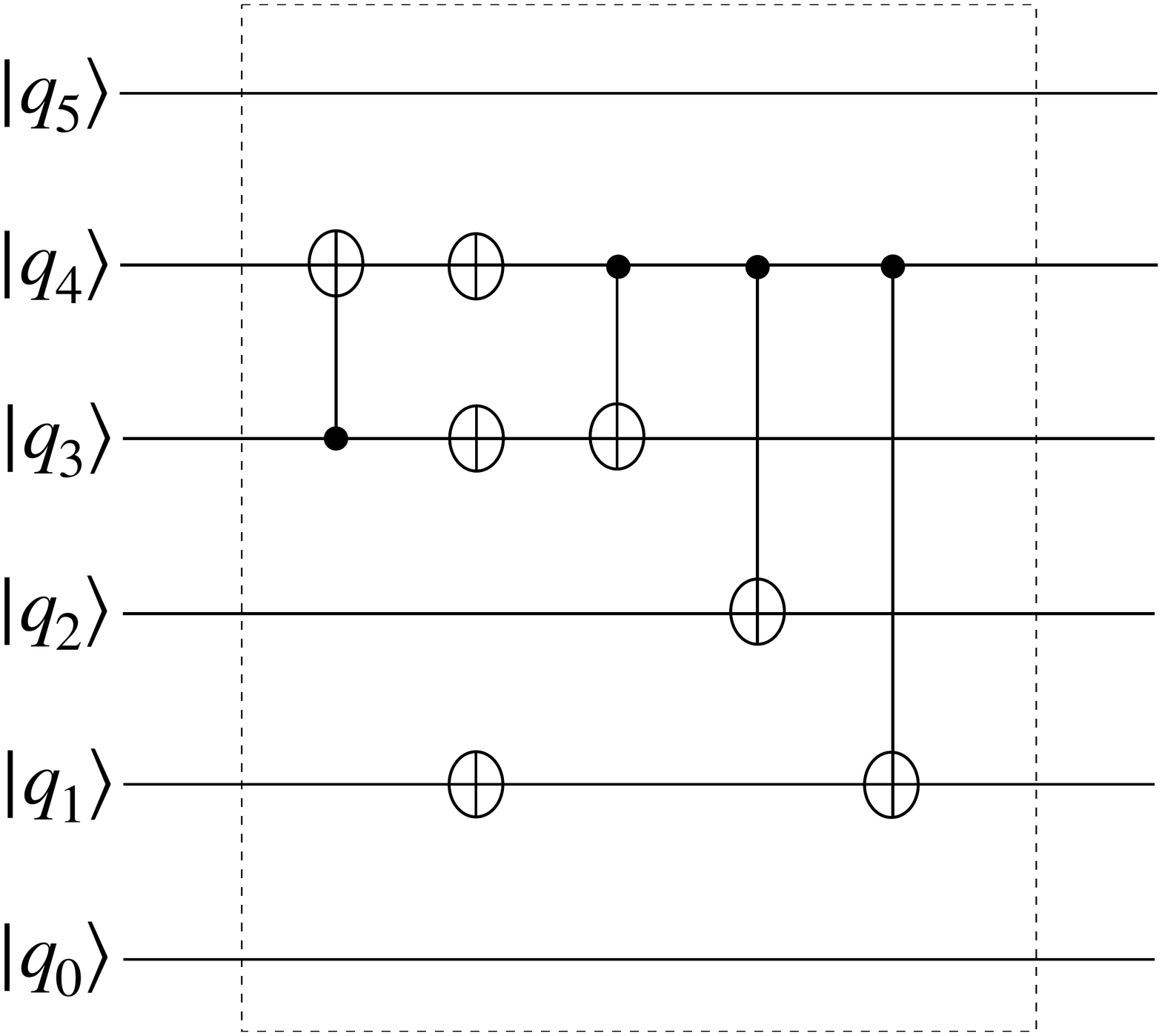}
		}
		\\	
	\subfloat[]{
		\label{circuit_35_4_1}
		\includegraphics[height = 1.2 in]{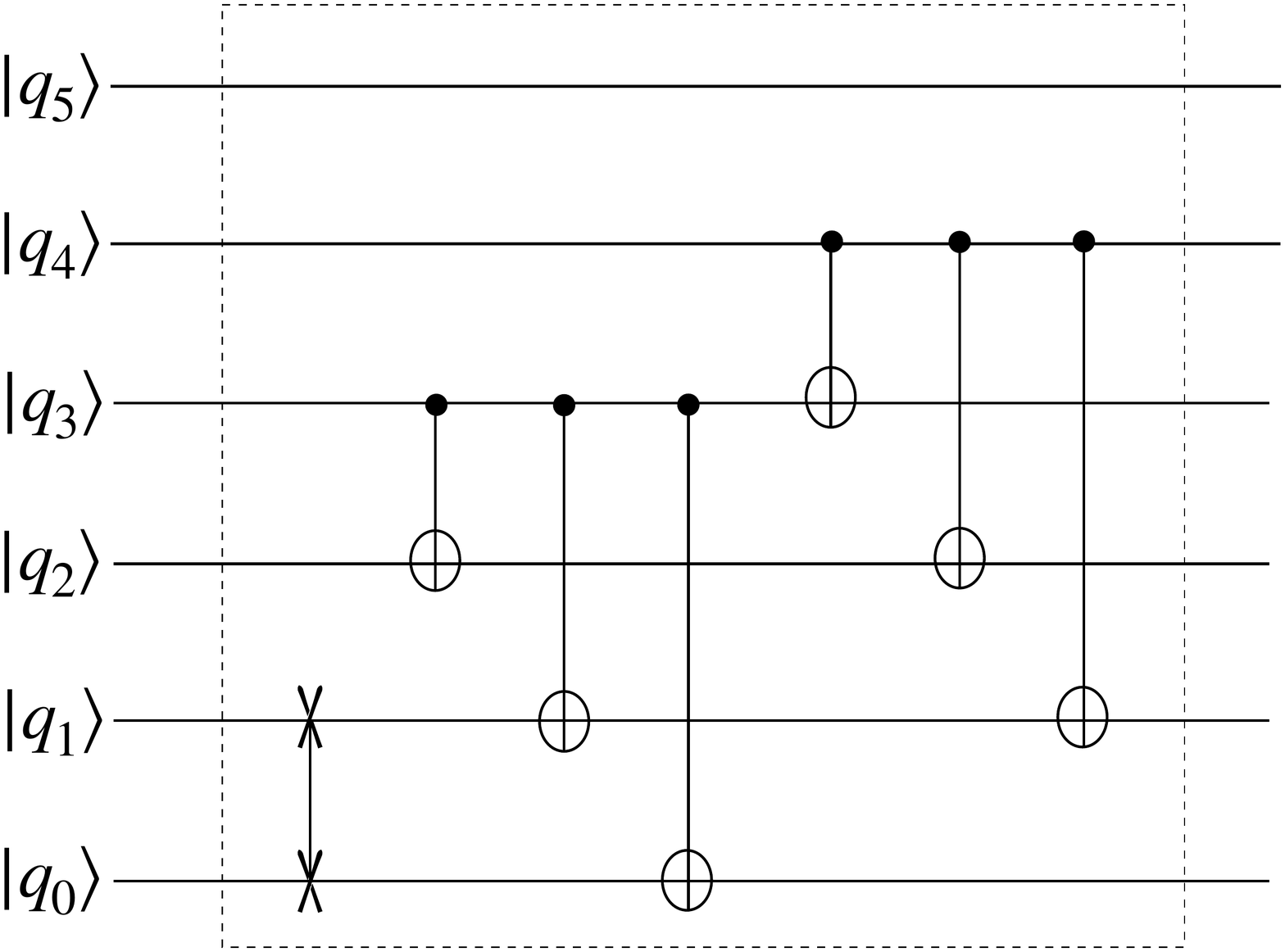}
		}	
	\caption{Modular exponentiation circuits for $N=35$. (a) $4^4 \,  \text{mod} \, 35$, (b) $4^2 \,  \text{mod} \,  35$, (c) $4^1 \,  \text{mod} \,  35$ .}
\label{mod_exp_35}	
\end{figure}

\end{document}